\documentclass[twocolumn,prb,superscriptaddress]{revtex4-2}

\usepackage{amsmath,amsfonts,amssymb}
\usepackage{graphicx,color,xcolor}
\usepackage{hyperref}
\usepackage{epstopdf}
\usepackage{float}
\usepackage{multirow}
\usepackage{hhline}
\usepackage{ulem}
\usepackage{mathtools}
\usepackage{siunitx}
\usepackage[caption=false]{subfig}
\usepackage{soul}
\hypersetup{hidelinks,colorlinks=true,
    linkcolor=blue,
    filecolor=magenta,      
    urlcolor=black,}

\newcommand{\ket}[1]{\left|#1\right\rangle}
\newcommand{\bra}[1]{\left\langle#1\right|}

\makeatletter

\@ifundefined{textcolor}{}
{%
 \definecolor{BLACK}{gray}{0}
 \definecolor{WHITE}{gray}{1}
 \definecolor{RED}{rgb}{1,0,0}
 \definecolor{GREEN}{rgb}{0,1,0}
 \definecolor{BLUE}{rgb}{0,0,1}
 \definecolor{CYAN}{cmyk}{1,0,0,0}
 \definecolor{MAGENTA}{cmyk}{0,1,0,0}
 \definecolor{YELLOW}{cmyk}{0,0,1,0}
}

\makeatother

\begin{document}

\title{Autonomous error correction of a single logical qubit using two transmons}

\author{Ziqian Li$^\ddag$}
\affiliation{James Franck Institute, University of Chicago, Chicago, Illinois 60637, USA}
\affiliation{Department of Physics, University of Chicago, Chicago, Illinois 60637, USA}
\affiliation{Department of Applied Physics, Stanford University, Stanford, California 94305, USA}
\altaffiliation{These authors contributed equally to this work}

\author{Tanay Roy$^\ddag$}
\email{Present address: Superconducting Quantum Materials and Systems Center, Fermi National Accelerator Laboratory (FNAL), Batavia, IL 60510, USA}
\affiliation{James Franck Institute, University of Chicago, Chicago, Illinois 60637, USA}
\affiliation{Department of Physics, University of Chicago, Chicago, Illinois 60637, USA}
\altaffiliation{These authors contributed equally to this work}

\author{David Rodr{\'i}guez P{\'e}rez}
\affiliation{Department of Physics, Colorado School of Mines, Golden, Colorado 80401, USA}

\author{Kan-Heng Lee}
\altaffiliation{Present address: Advanced Quantum Testbed, Lawrence Berkeley National Laboratory, Berkeley, CA 94720, USA}
\affiliation{James Franck Institute, University of Chicago, Chicago, Illinois 60637, USA}
\affiliation{Department of Physics, University of Chicago, Chicago, Illinois 60637, USA}

\author{Eliot Kapit}
\affiliation{Department of Physics, Colorado School of Mines, Golden, Colorado 80401, USA}

\author{David I. Schuster}
\affiliation{James Franck Institute, University of Chicago, Chicago, Illinois 60637, USA}
\affiliation{Department of Physics, University of Chicago, Chicago, Illinois 60637, USA}
\affiliation{Pritzker School of Molecular Engineering, University of Chicago, Chicago, Illinois 60637, USA}
\affiliation{Department of Applied Physics, Stanford University, Stanford, California 94305, USA}

\date{\today}

\begin{abstract}
Large-scale quantum computers will inevitably need quantum error correction to protect information against decoherence. Traditional error correction typically requires many qubits, along with high-efficiency error syndrome measurement and real-time feedback. Autonomous quantum error correction (AQEC) instead uses steady-state bath engineering to perform the correction in a hardware-efficient manner. We realize an AQEC scheme, implemented with only two transmon qubits in a 2D scalable architecture, that actively corrects single-photon loss and passively suppresses low-frequency dephasing using six microwave drives. Compared to uncorrected encoding, factors of 2.0, 5.1, and 1.4 improvements are experimentally witnessed for the logical zero, one, and superposition states. Our results show the potential of implementing hardware-efficient AQEC to enhance the reliability of a transmon-based quantum information processor.
\end{abstract}

\maketitle

\section{Introduction}

Quantum error correction (QEC) is critical for performing long computations involving many qubits, such as Shor's~\cite{shor1997} or quantum chemistry algorithms~\cite{chem2005}. Errors accumulating in the quantum system can be regarded as entropy or heat entering the system. In this context, the standard measurement and feedback-based QEC methods can be thought of us creating a ``Maxwell Demon" keeping the system cold. These methods typically require many qubits and complex control hardware and have been demonstrated approaching the fault tolerance threshold~\cite{Krinner2022, Abobeih2022, Bluvstein2022, Egan2021, anderson2021, Erhard2021, Cramer2016, Kelly2015, Waldherr2014, philipp2011}. When cooling atoms, rather than using measurement-based feedback, typically laser cooling is used. In laser cooling, the measurement and feedback are effectively encoded in the internal level structure and clever choice of laser drives.  Along these lines, it is possible to perform autonomous quantum error correction (AQEC) where rather than measurements and gates, the system is ``cooled'' via an appropriate set of drives and couplings to engineered thermal reservoirs~\cite{Verstraete2009}. Like laser cooling, AQEC can dramatically simplify the quantum and classical hardware and control required. Both autonomous and feedback-based QEC are more challenging than simply cooling because they require that the cooling process preserves the logical manifold of the system.

AQEC has received growing attention in theoretical proposals~\cite{wang2022, Albert_2019, eliot2018, Lihm2018, eliot2016, Joachim2014, Mirrahimi_2014, Zaki2013, Mohan2005}. In addition to the usual QEC conditions~\cite{KL2000}, AQEC requires that the error-correction operations must commute with the system Hamiltonian at all times. This makes AQEC most appropriate for hardware efficient~\cite{Gertler2021, Grimm2020, Campagne-Ibarcq2020, Ma2020, Hu2019, Ofek2016} systems with constrained error syndromes. Thus far all demonstrations, have encoded the logical qubits into 3D superconducting cavities using an ancilla qubit as a control~\cite{Gertler2021, Ma2020}. 

In this report, we experimentally realize AQEC in a pure transmon-based~\cite{koch2007transmon} system using scalable on-chip circuit structures. We propose a new AQEC protocol, called the Star code, which simplifies the original very small logical qubit (VSLQ) proposal~\cite{eliot2018, eliot2016} and does not require four-photon drive terms. We develop a coherence-preserving two-transmon coupler that can parametrically generate all interactions needed for the protocol. With AQEC turned on, the logical states show higher coherence times than the uncorrected case. The structure of the paper is as follows. First, we explain the logical encoding and Hamiltonian construction of the Star code. Then we experimentally calibrate each of the parametric processes used in the code. Finally, we prepare the logical states and characterize the coherence improvement.

\section{Theory}  
\begin{figure}[t]
    \centering
    \includegraphics[width=\columnwidth]{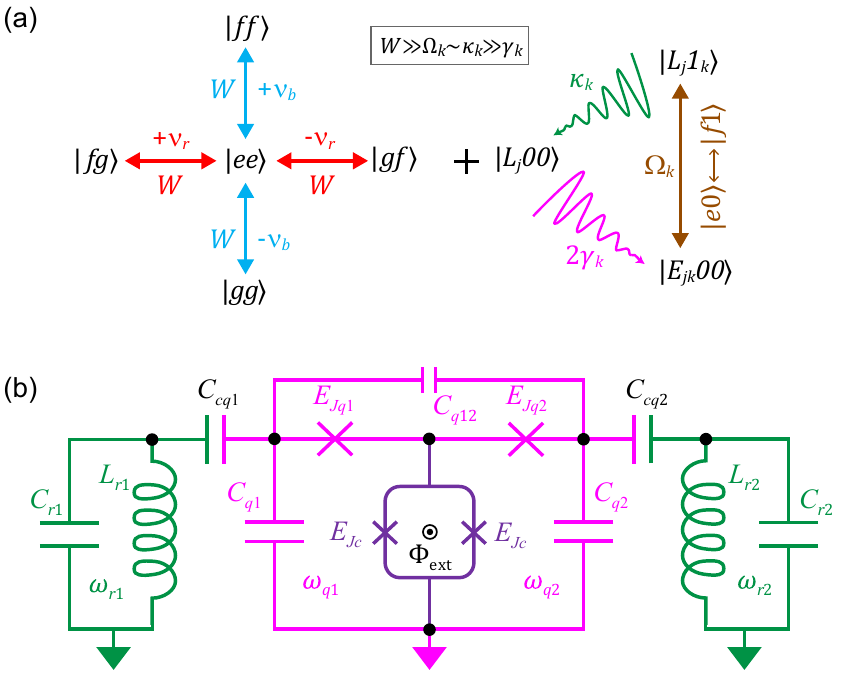}
    \caption{Schematic of Star code and circuit implementation. (a) Illustration of the autonomous error-correction scheme. The protocol requires simultaneous application of two QQ blue sidebands ($\ket{ee}\leftrightarrow\ket{gg}$ and $\ket{ee}\leftrightarrow\ket{ff}$), two QQ red sidebands ($\ket{ee}\leftrightarrow\ket{fg}$ and $\ket{ee}\leftrightarrow\ket{gf}$), and two QR error correcting sidebands ($\ket{e0}\leftrightarrow\ket{f1}$). All six drives are always-on. The red and blue QQ sidebands have nominally equal rates $W$ with equal and opposite detunings from the on-resonance values. The right part describes the AQEC cycle when a single-photon-loss event occurs. Logical state $\ket{L_j00}$ loses a photon from transmon $Q_{k}$ at rate $2\gamma_{k}$ and becomes the error state $\ket{E_{jk}00}$. QR error correcting sidebands bring the state at rate $\Omega_{k}$ to $\ket{L_j1_k}$ with one photon populating $R_{k}$. $R_{k}$'s photon decays quickly (at a rate $\kappa_k$) and recover the original logical state. (b) Circuit diagram for AQEC implementation. The device consists of two transmons, two resonators, and an inductive coupler.}
    \centering
    \label{fig:theory}
\end{figure}

The Star code encodes a logical qubit using two orthogonal states in a nine-dimensional (two-qutrit) Hilbert space as $\ket{L_{0}}=(\ket{gf}-\ket{fg})/\sqrt{2}$ (logical ``zero"), and $\ket{L_{1}}=(\ket{gg}-\ket{ff})/\sqrt{2}$ (logical ``one") where $\ket{g},\ket{e}$, and $\ket{f}$ represent the lowest three energy levels of a transmon. The error states after a single photon-loss (one transmon in $\ket{e}$) are orthogonal to the logical space and to each other. Further, both logical states have an equal expected photon number so that photon loss does not reveal information about the state it was emitted from.  We engineer a parent Hamiltonian for the logical states through $\ket{gf}\bra{fg}$ and $\ket{gg}\bra{ff}$ parametric processes. These processes are all implemented by driving through $\ket{ee}$ as an intermediate state, producing the star topology in Hilbert space that gives the code its name (see Fig.~\ref{fig:theory}a). Using an intermediate state allows these to be achieved using only 2-photon drives (QQ sidebands) rather than the higher-order 4-photon processes required by the VSLQ~\cite{eliot2016}. Despite both sets of drives going through $\ket{ee}$, with careful tuning of the drive $W$, the logical states can be made dark with respect to $\ket{ee}$ by detuning the $\ket{L_0}$ ($\ket{L_1}$) sidebands by $\pm \nu_r$ ($\pm \nu_b$). When all of these processes are simultaneously applied, the two-transmon Hamiltonian in the logical-static frame (see Appendix~\ref{section:frame} for derivation) is
\begin{align}
    \tilde{H}_{QQ}=&\frac{W}{2}\left(\ket{ee}\bra{gf}e^{2\pi i\nu_r t}+\ket{ee}\bra{fg}e^{2\pi i\nu_r t}\right.\nonumber\\
&\left.+\ket{ee}\bra{gg}e^{2\pi i\nu_b t}+\ket{ee}\bra{ff}e^{2\pi i\nu_b t}\right)+h.c.
    \label{eq:QQ_sideband}
\end{align} 

Each transmon $Q_j$ is coupled to a lossy resonator $R_j$, which acts as the cold reservoir for entropy dumping. A single-photon loss, the dominant source of error in the system, populates the $\ket{e}$ level, triggering autonomous correction enabled by two transmon-resonator (QR) error correcting sidebands $\ket{e0}_j\leftrightarrow\ket{f1}_j, j=1,2$ (right part of Fig.~\ref{fig:theory}a). These sidebands are applied resonantly at rates $\Omega_{j}$ to the system, adding $\tilde{H}_{QRj}$ to the system Hamiltonian $H_{\rm static}$,
\begin{align}
    \tilde{H}_{QR1}=&\frac{\Omega_1}{2}a^{\dagger}_{r1}\left(\ket{fg}\bra{eg}+\ket{ff}\bra{ef}\right)\otimes I_{4}+h.c. , \nonumber \\
\tilde{H}_{QR2}=&\frac{\Omega_2}{2}a^{\dagger}_{r2}\left(\ket{gf}\bra{ge}+\ket{ff}\bra{fe
}\right)\otimes I_{4}+h.c.,
    \label{eq:QR_sideband}\\
    \tilde{H}_{\rm static}=&\tilde{H}_{QQ}\otimes I_{4}+\sum_{j=1,2}\tilde{H}_{QRj}+H_c.
\label{eq:fully_rotated}
\end{align}
Here $a_{rj}$ is the annihilation operator for the $j$-th resonator, and $\alpha_{j}$ is the anharmonicity of $j$-th transmon. $H_c$ contains the diagonal terms from frame transformation. We label the full state as $\ket{Q_{1}Q_{2}R_{1}R_{2}}$. We keep the lowest two levels for each resonator, and $I_{n}$ is the $n\times n$ identity matrix.

The Star code can correct the loss of a single photon from one of the qubits. Suppose $Q_{1}$ loses a photon at rate $2\gamma_{1}$, where $\gamma_1$ is the $\ket{e}\rightarrow \ket{g}$ decay rate. The logical $\ket{L_000}$, consequently, becomes the error state $\ket{E_{01}00}=\ket{eg00}$ with energy $-\frac{\alpha_{1}}{2}$. When $W\gg\Omega_{j}$, $\tilde{H}_{QRj}$ is a perturbation and only drives the transition $\ket{E_{01}00}\leftrightarrow\ket{L_{0}10}$ (See Fig.~\ref{fig:theory} (a)). Assuming the resonator's decay rate $\kappa_1\gg\gamma_{1}$, this oscillation quickly damps back to the original logical state $\ket{L_{0}00}$ with no extra phase accumulated, and completes the correction cycle. The correction procedure for $\ket{L_{1}}$ is similar through an independent path. The logical superposition state preserves relative phases since the QR sidebands do not distinguish the correction path. Such a two-step logical refilling rate can be approximated with Fermi's golden rule $\Gamma_{Rj} \simeq \frac{\Omega_{j}^{2}\kappa_{j}}{\Omega_{j}^{2}+2\kappa_{j}^{2}}$~\cite{eliot2014}. Apart from providing protection against single-photon loss, the star code also provides suppression to $1/f$ dephasing error~\cite{eliot2016,kapit2017review}. The continuous QQ drives create an energy gap between the logical manifold and all other states, suppressing low-frequency noise. Theoretical lifetime improvement of logical states is further discussed in Ref.~\cite{VStar2023}.

We realize this protocol using the circuit shown in Fig.~\ref{fig:theory}b. The key component is the inductive coupler based on the design in Ref.~\cite{Yao2017} that enables the realization of fast parametric interactions. Two transmons $Q_{1}$ and $Q_{2}$ serve as the qutrits and share a common path to ground. This path is interrupted by a Superconducting Quantum Interference Device (SQUID) loop. The SQUID functions as a tunable inductor with external DC and RF magnetic fields threaded for activating the QQ sidebands. Each transmon is capacitively coupled to a lossy resonator serving both as the readout and cold reservoir. QR sidebands can be performed by sending a charge drive at the half transition frequency to the transmon~\cite{wallraff2007sidband}. Full circuit quantization is shown in Appendix~\ref{app:sideband rate}.

\section{Experimental results}

\subsection{Device implementation and sideband calibration}
\begin{figure}[t]
    \centering
    \includegraphics[width=\columnwidth]{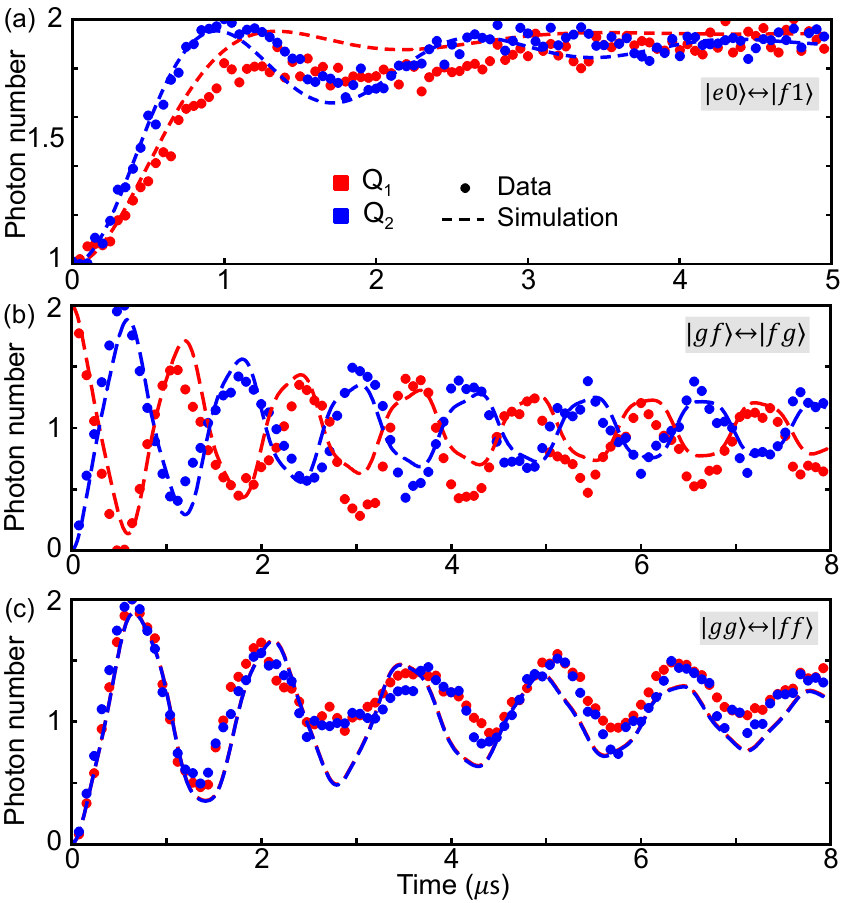}
    \caption{Different parametric oscillations. (a) Error correcting QR sidebands $\ket{e0}\leftrightarrow\ket{f1}$ applied separately at rates $\Omega_{1}=\SI{0.49}{\mega\hertz}$ and $\Omega_{2}=\SI{0.59}{\mega\hertz}$ to the transmon-resonator pairs with $\ket{e}$ as initial states. Effective transitions (b) $\ket{gf}\leftrightarrow\ket{fg}$ and (c) $\ket{gg}\leftrightarrow\ket{ff}$ are measured when all QQ and QR sidebands are simultaneously turned on. Extracted sideband rates and detunings from simulation are $W_{r}=1.45$ MHz, $W_{b}=1.25$ MHz, $\nu_{r}=0.8$ MHz, $\nu_{b}=-0.9$ MHz, $\Omega_{1}=\Omega_{2}=0.39$ MHz. Oscillation distortions are quantitatively matched in the lab frame simulations.}
    \centering
    \label{fig:6sideband}
\end{figure}

In this section, we will characterize the individual qubits and realize the required sidebands to create and correct the logical states. We adjust the DC flux point to minimize the Cross-Kerr coupling between transmons which can dephase the logical superposition states (See Appendix~\ref{section:ZZ} for further discussion). The measured Cross-Kerr couplings are all lower than $\SI{320}{\kilo\hertz}$ while maintaining Ramsey dephasing times $\ T_{R_{ge}}=15.2(9.8)\ \mu$s\ with relaxation time $T_{1ge}=24.3(9.1)\ \mu$s, for $Q_{1}(Q_{2})$ (See Appendix~\ref{app:device parameters}). 

To calibrate the QR sidebands for selective photon pumping, we initialize the system in $\ket{eg00}$ and apply a continuous charge drive at frequency $(\omega_{r1}+\omega_{q1}+\alpha_{1})/2$ to activate a 2-photon $\ket{e0}\leftrightarrow\ket{f1}$ transition between $Q_1$ and $R_1$ at a rate of $\SI{0.49}{\mega\hertz}$. The system achieves a steady state $\ket{fg00}$ within $\SI{3}{\micro\second}$ as shown by red points in Fig.~\ref{fig:6sideband}(a). Similarly, a $\SI{0.59}{\mega\hertz}$ QR2 drive takes $\ket{ge00}$ to $\ket{gf00}$ in a similar time (blue points in Fig.~\ref{fig:6sideband}(a)). The decay of transmon reduces the final average photon number slightly below 2.

We achieve at least $20$ MHz QQ red sidebands ($\left(\ket{j,k}\leftrightarrow\ket{j+1,k-1}\right)$) and $5$ MHz QQ blue sidebands ($\left(\ket{j,k}\leftrightarrow\ket{j+1,k+1}\right)$) separately at the operating point, demonstrating a fast, coherence-preserved two-qutrit coupler with suppressed $ZZ$ interaction. Blue sidebands have a slower rate limited by stray signals from higher flux modulation frequencies (See discussion in Appendix~\ref{app:readout problem}). All possible sidebands realized in this coupler are shown in Appendix~\ref{app:sidebandexp}.

By driving all six sidebands, the core effective 4-photon processes, $\ket{fg}\leftrightarrow\ket{gf}$ and $\ket{gg}\leftrightarrow\ket{ff}$ and the error-correcting QR drives can be realized simultaneously. In practice, the QQ red and blue sideband rates ($W_{r}=\SI{1.45}{\mega\hertz}$ and $W_{b}=\SI{1.25}{\mega\hertz}$) are slightly different. When applying all sidebands, we choose a smaller $W$, because the coupler was found to heat and shift the readout resonator when driven at larger rates making tomographic reconstruction inaccurate. We choose almost opposite detunings ($\nu_{r}=\SI{0.8}{\mega\hertz}$, $\nu_{b}=\SI{-0.9}{\mega\hertz}$) for larger energy separation of the eigenstates and better error correction performance. Both QR sidebands are turned on at rates $\Omega_1=\Omega_2=\SI{0.39}{\mega\hertz}$. Fig.~\ref{fig:6sideband}(b) shows the evolution when the initial state is $\ket{gf}$. The average photon number of $Q_{1}$ (in red) and $Q_{2}$ (in blue) are read out simultaneously, and the oscillation between 0 and 2 forms an effective 4-photon red sideband. Note that this effective swap process is slightly different from the direct $\ket{fg} \leftrightarrow \ket{gf}$ transition as the population in $\ket{ee}$ will appear intermediately when the initial state has overlap with the eigenstates that have $\ket{ee}$ component. Under this condition, $\ket{ee}$ is no longer the dark state of the mixed QQ sidebands. Oscillation damping originates from the detuning-induced slow interference and decoherence of the qutrit subspace, and this distortion is captured by the simulation as well. Similarly, by choosing the initial state as $\ket{gg}$, the effective four-photon blue sideband $\ket{gg}\leftrightarrow\ket{ff}$ can be observed in Fig.~\ref{fig:6sideband}(c). 

\begin{figure*}[t]
    \centering
    \includegraphics[width=\textwidth]{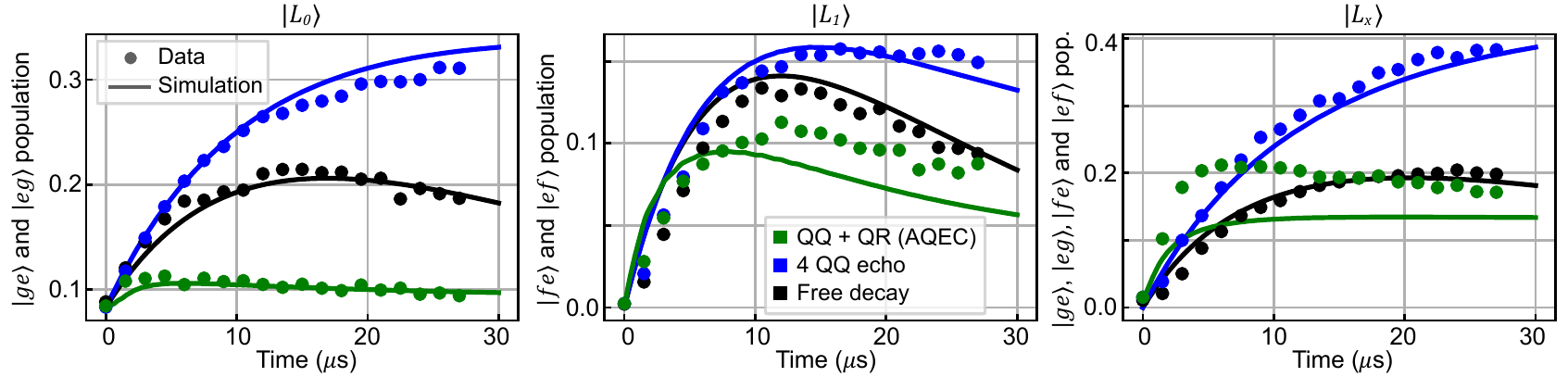}
    \caption{Error population under different conditions. Black, blue, and green points represent tomographic measurement results under free decay, 4 QQ echo, and full AQEC. The y-axes represent the combined population of error states for initial states $\ket{L_0}, \ \ket{L_1}$, and $\ket{L_x}$. Population accumulates at the error states in the free decay case, enhanced in the 4 QQ echo case, and corrected with AQEC drive on. The experimental data is explained with master equation simulations. Detailed simulation parameters are shown in the Appendix~\ref{app:simulation}.}
    \centering
    \label{fig:correctable}
\end{figure*}

\subsection{Error Correction Performance}

The logical state initialization requires sequential application of multiple single-qutrit and two-{qutrit rotations. For $\ket{L_0}$ and $\ket{L_1}$, QQ red and blue sidebands are used to generate entanglement, and for $\ket{L_x}=(\ket{L_0}+\ket{L_1})/\sqrt{2}=(\ket{g}+\ket{f})(\ket{g}-\ket{f})/2$, only single qutrit rotations are required. The preparation times for initial states are separately $\SI{313}{\nano\second}$, $\SI{142}{\nano\second}$, and $\SI{282}{\nano\second}$ for $\ket{L_0}$, $\ket{L_1}$ and $\ket{L_x}$. The detailed preparation circuit is discussed in Appendix~\ref{app:calibration}. We perform full two-qutrit state tomography~\cite{Wallraff2010qutrit_tomo, Tanay2021tomography} and obtain initial state fidelities of $88.1\%$, $89.1\%$ and $88.7\%$ for the three states respectively. The tomography sequences and density matrix reconstruction are shown in Appendix~\ref{app:tomography}.

We characterize the performance of the Star code by comparing three different cases --- free decay, QQ sideband spin-locking (4 QQ echo), and full AQEC. For free decay, we do not apply any drive after the state preparation. For the 4 QQ echo case, we turn on the QQ sidebands $\ket{ee}\leftrightarrow\{\ket{gf},\ket{fg},\ket{gg},\ket{ff}\}$ with a similar rate-detuning configuration as shown in Fig.~\ref{fig:theory}a ($W_r=\SI{1.0}{\mega\hertz}, W_b=\SI{1.7}{\mega\hertz}, \nu_{r}=\SI{1.5}{\mega\hertz}, \nu_{b}=\SI{0.0}{\mega\hertz}$). This case shows coherence improvement from spin-locking. The full AQEC ($W_{r}=1.45$ MHz, $W_{b}=1.25$ MHz, $\nu_{r}=0.8$ MHz, $\nu_{b}=-0.9$ MHz, $\Omega_{1}=\Omega_{2}=0.39$ MHz) demonstrates further improvement from photon-loss correction. We plot the density matrices of the logical states after preparation and after $\SI{9}{\micro\second}$ in Appendix~\ref{app:tomography} for reference.

To demonstrate that our protocol corrects single-photon loss error, in Fig.~\ref{fig:correctable}, we plot the combined population of error states as a function of time for all three cases. The error populations are computed through the expectation values of $\varepsilon_0 = \ket{ge}\bra{ge}+\ket{eg}\bra{eg}$ for $\ket{L_0}$, $\varepsilon_1 = \ket{ef}\bra{ef}+\ket{fe}\bra{fe}$ for $\ket{L_1}$, and $\varepsilon_0 + \varepsilon_1$ for $\ket{L_x}$ corresponding to the states after single-photon loss. We extract the error population from the density matrices reconstructed with full two-qutrit state tomography at each time point up to $\SI{27}{\micro\second}$ using the Maximum Likelihood Estimation (MLE) from 5000 measurements for each state. This is a direct demonstration of the AQEC's effectiveness, as it measures the error state population designed to correct by the protocol. Compared to the free decay cases (black dots), turning on the AQEC clearly corrects photon loss and suppresses the error rate below the free decay cases (green dots). The error rates for all three logical states increase in the 4 QQ echo case (blue dots), as enhanced qutrit decay rates in the presence of sideband} can lead to extra photon loss. The solid lines represent rotating frame simulations and are in agreement with the experimental data.

\begin{figure}[t]
    \includegraphics{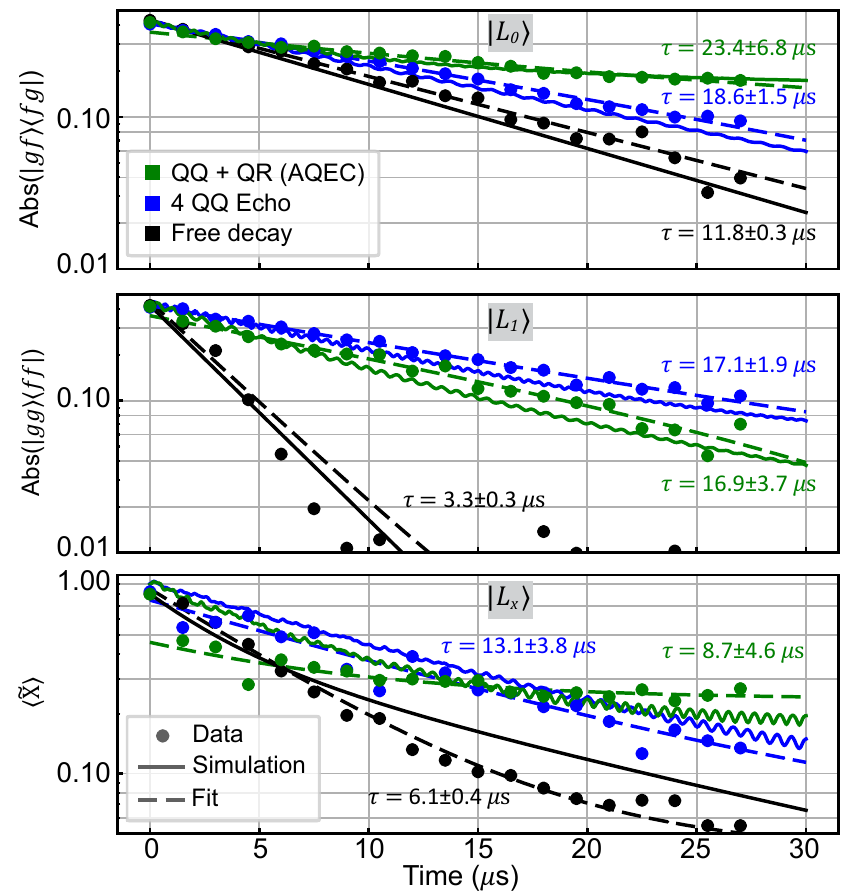}
    \caption{Coherence improvement. Black, blue, and green circles are experimentally obtained expectation values for the relevant operators representing coherence at a given time. The expectation values are extracted from the tomographic reconstruction of states with 5000 repeated measurements. The improvement with AQEC turned on is explained by the master equation simulation. All traces are fitted to the exponential decay curve $A \exp(-t/\tau)+C$. The error bars (one standard deviation) for $\tau$ are obtained from the fitting$^*$. The fast transition period (first $\SI{1.5}{\micro\second}\sim{\Omega_j^{-1}}$ in the AQEC case is not included in the fitting for a better representation of logical coherence.\\
    $^{*}$Large error comes from treating $C$ as a free variable in the fitting.}
    \centering
    \label{fig:improvement}
\end{figure}

In addition to correcting photon loss, it is also important to characterize how well the AQEC protocol preserves the coherence of the logical states. To quantify the coherence, we plot the decay of the most sensitive off-diagonal matrix element for each logical state. Fitting the data to the exponential decays for $\ket{L_0}$ and $\ket{L_1}$~\cite{DavidR2021}, the logical states' coherence are improved from $\SI{11.8}{\micro\second}$ ($\ket{L_0}$) and $\SI{3.3}{\micro\second}$ ($\ket{L_1}$) in the free decay cases, to $\SI{18.6}{\micro\second}$ and $\SI{17.1}{\micro\second}$ in the four QQ echo cases, and up to $\SI{23.4}{\micro\second}$ and $\SI{16.9}{\micro\second}$ in the error correction cases (see Fig.~\ref{fig:improvement}(a), (b)). This demonstrates a factor of $2.0$ and $5.1$ improvement in logical state coherence against the free decay case. We use the operator $\tilde{X}=\left(\ket{gg}+\ket{fg}\right)\left(\bra{gf}+\bra{ff}\right)/2+h.c.$ for showing $\ket{L_x}$'s coherence. $\tilde{X}$ is the projection of the error-transparent logical X rotation $\left(I+X_{1}\right)X_{2}/2$ to the $\{\ket{g},\ket{f}\}$ subspace, with $X_{j}=(a^{\dagger}_{qj}a^{\dagger}_{qj}+a_{qj}a_{qj})/\sqrt{2}$~\cite{eliot2018}. In Fig.~\ref{fig:improvement}, the solid and dashed lines represent rotating-frame simulations and fit, matching qualitatively to the experimental data. Our results demonstrate a factor of $1.4$ improvement in $\ket{L_x}$'s lifetime.

The large difference in free-decay coherence times between $\ket{L_0}$ and $\ket{L_1}$ originates from the low-frequency dephasing noise on $\Phi_{\rm DC}$ through the flux line. It causes a shift in both transmons' frequencies in the same direction, which $\ket{L_1}$ is sensitive to but $\ket{L_0}$ is not. The passive echo protection from the Star code drives suppresses this; consequently, in the 4 QQ echo case both logical states have similar coherence time.

The AQEC performance is primarily limited by three factors in our experiment. The most important fact is that the QQ sideband rates $W_r$ and $W_b$ are well below their ideal values. Stronger drives would further suppress phase noise (lifetimes in the 4 QQ echo experiment are well below $2T_{1}$, indicating room for improvement), and the increased energy separation would also allow us to use stronger QR drives, correcting photon loss more quickly. Although the coupler supports $\SI{9}{\mega\hertz}$ QQ sidebands for short periods, when $W_{b}$ goes beyond $\SI{5}{\mega\hertz}$ the readout resonator frequency starts to shift, introducing systematic measurement distortion (See Appendix~\ref{app:readout problem} for details). This problem worsens with all six tones applied and we stay well below this limit to ensure reliable tomography results. The second limit is the $ZZ$ coupling between the transmons, an extra dephasing channel for superposition states (see Appendix~\ref{section:ZZ} for details). Our coupler is operated at the minimum $ZZ$ flux bias of the coupler to minimize the effect. It could be further mitigated by stronger QR sidebands enabling faster error correction, or through additional off-resonant QQ drive terms to dynamically cancel it. The third limit comes from heating and physical coherence drop when sidebands are turned on. The average photon number in the readout increases from $<0.01$ (free decay and 4 QQ echo cases) to $0.03$ (AQEC case), and the photon-excitation event in the transmon is a non-correctable error source (see Appendix~\ref{app:simulation}). This explains why a clear reduction in correctable error rate does not result in a comparable increase in the logical lifetime. Further improvement can thus come from two paths---improving isolation between control signals or improving physical qubit coherence so that weaker drives can be more effective. Other limits are in the order of ms as shown in Appendix~\ref{app:simulation} and do not affect our results considerably.

\section{Conclusion and Outlook}

We have experimentally demonstrated a hardware-efficient AQEC code, the Star code, that requires only two transmon-resonator pairs and a linear coupler to perform the second-order transitions. Three levels per transmon are used to store information, with the middle level capturing photon loss error, and entropy is dumped to the resonator autonomously through the always-on cooling sidebands. Inter-transmon parametric drives are applied to the coherence-preserving coupler for separating the Star code logical space. We demonstrate AQEC's effectiveness in a pure transmon system that is free from 3D cavities, compared to previous AQEC demonstrations in the bosonic system. Our system is entirely constructed from scalable components and fundamentally avoids the need for fast and accurate error detection and feedback error correction pulses. The Star code can be a self-corrected building block for the surface code~\cite{Krinner2022, youwei2022} to further correct higher-order errors when scaled up.

Future work will include realizing proposed error-transparent single-qubit and multi-qubit gates~\cite{eliot2018}.  The Star code can also be implemented in other platforms that have full control of multiple anharmonic three-level systems. 



\section{ACKNOWLEDGEMENT}
This work was supported by AFOSR Grant No.
FA9550-19-1-0399 and ARO Grant No. W911NF-17-S0001. Devices are fabricated in the Pritzker Nanofabrication Facility at the University of Chicago, which receives
support from Soft and Hybrid Nanotechnology Experimental (SHyNE) Resource (NSF ECCS-1542205), a node
of the National Science Foundation’s National Nanotechnology Coordinated Infrastructure. This work also made use of the shared facilities at the University of Chicago Materials Research Science and Engineering Center, supported by the National Science Foundation under award number DMR-2011854. EK's research was additionally supported by NSF Grant No. PHY-1653820.

\clearpage

\appendix

\section{Device Parameters}
\label{app:device parameters}
\begin{table}[h]
		\begin{tabular}{c c c c}
		    \hline
		    \hline
		      Transition & $T_{1} \left(\mu \text{s}\right)$ & $T_{R} \left(\mu \text{s}\right)$ & $T_{echo} \left(\mu \text{s}\right)$ \\  \hline
		    $Q_{1} \ket{e}\rightarrow\ket{g} $ & $24.3$ & $15.2$ & $24.6$ \\  
		    $Q_{2} \ket{e}\rightarrow\ket{g}$ & $9.1$ & $9.8$ & $14.3$ \\  
		    $Q_{1} \ket{f}\rightarrow\ket{e} $ & $27.1$ & $16.7$ & $29.3$ \\  
		    $Q_{2} \ket{f}\rightarrow\ket{e}$ & $26.7$ & $20.1$ & $34.3$\\  \hline
		    $R_{1} \ket{1}\rightarrow\ket{0} $ & $0.3$ &  &  \\  
		    $R_{2} \ket{1}\rightarrow\ket{0}$ & $0.3$ &  & \\  \hline
		\end{tabular}
		\caption{Device coherence parameters.}
		\label{table:qutrit_coherence}
\end{table}
 \begin{table}[b]
  \begin{tabular}{c c c}
		    \hline
		    \hline
		     Parameter & Symbol & Value/$2\pi$ \\  \hline
		     $Q_{1}$ ge frequency & $\omega_{q1}$ & $\SI{3.2049}{\giga\hertz}$  \\  
		     $Q_{2}$ ge frequency & $\omega_{q2}$ & $\SI{3.6625}{\giga\hertz}$  \\  
		     $Q_{1}$ anharmonicity & $\alpha_{1}$ & $\SI{-116.4}{\mega\hertz}$  \\  
		     $Q_{2}$ anharmonicity & $\alpha_{2}$ & $\SI{-159.6}{\mega\hertz}$  \\  \hline
		     $R_1$ frequency & $\omega_{r1}$ & $\SI{4.9946}{\giga\hertz}$  \\  
		     $R_2$ frequency & $\omega_{r2}$ & $\SI{5.4505}{\giga\hertz}$  \\ 
		     $R_1$ dispersive shift & $\chi_{1}$ & $\SI{-180}{\kilo\hertz}$ \\
		     $R_2$ dispersive shift & $\chi_{2}$ & $\SI{-330}{\kilo\hertz}$ \\		     \hline
		     $\left(E_{\ket{ee}}-E_{\ket{ge}}\right)-\left(E_{\ket{eg}}-E_{\ket{gg}}\right)$ & $ZZ_{ge}$ & $\SI{-261}{\kilo\hertz}$ \\ 
		     $\left(E_{\ket{fe}}-E_{\ket{ee}}\right)-\left(E_{\ket{fg}}-E_{\ket{eg}}\right)$ & $ZZ_{ef1}$ & $\SI{-130}{\kilo\hertz}$ \\ 
		     $\left(E_{\ket{ef}}-E_{\ket{ee}}\right)-\left(E_{\ket{gf}}-E_{\ket{ge}}\right)$ & $ZZ_{ef2}$ & $\SI{-301}{\kilo\hertz}$ \\ 
		     $\left(E_{\ket{ff}}-E_{\ket{ef}}\right)-\left(E_{\ket{fg}}-E_{\ket{eg}}\right)$ & $ZZ_{ff1}$ & $\SI{-171}{\kilo\hertz}$ \\ 
		     $\left(E_{\ket{ff}}-E_{\ket{fe}}\right)-\left(E_{\ket{gf}}-E_{\ket{ge}}\right)$ & $ZZ_{ff2}$ & $\SI{-289}{\kilo\hertz}$  \\ 
		     $\left(E_{\ket{ef}}-E_{\ket{gf}}\right)-\left(E_{\ket{eg}}-E_{\ket{gg}}\right)$ & $ZZ_{gf1}$ & $\SI{-619}{\kilo\hertz}$ \\ 
		     $\left(E_{\ket{fe}}-E_{\ket{fg}}\right)-\left(E_{\ket{ge}}-E_{\ket{gg}}\right)$ & $ZZ_{gf2}$ & $\SI{-464}{\kilo\hertz}$ \\ \hline
		     Coefficient of $n_{q1}n_{q2}$ & $J_{11}$ & $\SI{-312}{\kilo\hertz}$ \\ 
		     Coefficient of $n_{q1}^2n_{q2}$ & $J_{21}$ & $\SI{25}{\kilo\hertz}$ \\ 
		     Coefficient of $n_{q1}n_{q2}^2$ & $J_{12}$ & $\SI{-49}{\kilo\hertz}$ \\ 
		     Coefficient of $n_{q1}^2n_{q2}^2$ & $J_{22}$ & $\SI{-43}{\kilo\hertz}$ \\ \hline
		\end{tabular}
		\caption{Device frequencies without external drives.}
		\label{table:frequency}
\end{table}

Relevant coherence parameters and frequencies at the operating point (Coupler DC flux bias $\Phi_{\rm DC}=0.3795\Phi_{0}$) without external drives are listed in Table.~\ref{table:qutrit_coherence} and Table.~\ref{table:frequency}. The $ZZ$ coupling (dispersive shifts) between two-transmon energy levels are measured in the experiment through Ramsey fringe frequency difference (Appendix~\ref{app:calibration}), and the cross-Kerr couplings $J_{11}$, $J_{21}$, $J_{12}$, $J_{22}$ are calculated from the measurement results.
\begin{table}[t]
    \begin{tabular}{c c c c}
		    \hline
		    \hline
		     Capacitance & (fF) & Josephson Energy & (GHz) \\  \hline
		    $C_{q1}$ & $165.9$ & $E_{j1}$ & $12.4$ \\  
		    $C_{q2}$ & $123.4$ & $E_{j2}$ & $12.1$ \\  
		    $C_{c}$ & $178.3$ & $E_{jc}$ & $1106.0$ \\  
		    $C_{q12}$ & $2.0$ &  &  \\  \hline
	\end{tabular}
		\caption{Capacitances and Josephson energies used in the simulation. Capacitances are extracted through design geometry simulations in ANSYS Q3D, and Josephson energies are calculated from the room temperature resistances of identical test junctions on the same chip.}
	\label{table:qutrit_params}
\end{table}

\section{Star code frame transformation}
\label{section:frame}
We explicitly show the Hamiltonian for the Star code in different frames, with the sideband parameter set $\{W_r=W_b=W, \Omega_j=\Omega, \nu_r, \nu_b\}$. Without the external drives, the device can be described by the following Hamiltonian in the lab frame by keeping leading-order terms,
\begin{align}
    H = & \sum_{j=1}^2\left(\omega_{qj}n_{qj} + \dfrac{\alpha_j}{2}n_{qj}\left(n_{qj}-1\right) + \omega_{rj}n_{rj}+\chi_{j}n_{qj}n_{rj}\right)  \nonumber\\
    & + \sum_{j,k=1,2}J_{jk} (n_{q1})^j (n_{q2})^k.
\end{align}
Here $n_{qj} = a^\dagger_{qj}a_{qj}$ and $n_{rj}=a^{\dagger}_{rj}a_{rj}$ are the photon number operators for the $j$-th transmon and resonator respectively with $a_{qj (rj)}$ representing the annihilation operator for the $j$-th transmon (resonator). $\chi_{j}$ and $J_{jk}$ are the $ZZ$ coupling strength between the $j$-th QR pair and between two transmons. Inter-transmon $ZZ$ couplings are expanded to the second order for explaining shifts up to $\ket{f}$ level. Cross-Kerr couplings $J_{jk}$ are minimized by biasing DC flux at $\Phi_{\rm DC}=0.3795\Phi_0$ to suppress logical states' extra dephasing channel (see derivation in Appendix~\ref{section:ZZ}). 

Then we bring in external drives and ignore the static $ZZ$ couplings between QQ and QR, which can be reintroduced into the equation by shifting the diagonal energies. The \textit{lab-frame} Hamiltonian then reads
\begin{align}
H_{\rm lab} =& \sum_{j=1,2}\left(\omega_{qj}a_{qj}^{\dagger}a_{qj}+\frac{\alpha_{j}}{2}a_{qj}^{\dagger}a_{qj}^{\dagger}a_{qj}a_{qj}+\omega_{rj}a_{rj}^{\dagger}a_{rj}\right) \nonumber\\
&+H_{QQ}+H_{QR1}+H_{QR2} \label{eq:Hlab},\\
H_{QQ}=& A_{QQ}\left(t\right)\left(a_{q1}^{\dagger}+a_{q1}\right)\left(a_{q2}^{\dagger}+a_{q2}\right) \nonumber,\\
H_{QRj}=& A_{QRj}\left(t\right)\left(a_{qj}^{\dagger}+a_{qj}\right)\left(a_{rj}^{\dagger}+a_{rj}\right) \nonumber,\\
A_{QQ}\left(t\right) = & \frac{W}{\sqrt{2}}\cos{\left(\left(\omega_{q2}-\omega_{q1}-\alpha_{1}-\nu_{r}\right)t\right)} \nonumber\\
&+ \frac{W}{\sqrt{2}}\cos{\left(\left(\omega_{q2}-\omega_{q1}+\alpha_{2}+\nu_{r}\right)t\right)} \nonumber\\
&+ W\cos{\left(\left(\omega_{q1}+\omega_{q2}-\nu_{b}\right)t\right)} \nonumber\\
&+ \frac{W}{2}\cos{\left(\left(\omega_{q1}+\omega_{q2}+\alpha_{1}+\alpha_{2}+\nu_{b}\right)t\right)} \nonumber,\\
A_{QRj}\left(t\right) = & \frac{\Omega}{\sqrt{2}}\cos{\left(\left(\omega_{qj}+\omega_{rj}+\alpha_{j}\right)t\right)}.\nonumber 
\end{align}

Next, we move to the ``\textit{logical-static}" frame where all logical states have zero energy resulting in (we keep the lowest two levels for the resonators)
\begin{align}
\tilde{H}_{\rm static}=&-\frac{\alpha_{1}}{2}(P_{eg}+P_{ef})-\frac{\alpha_{2}}{2}(P_{ge}+P_{fe}) \nonumber\\
&+\tilde{H}_{QQ} \nonumber\\
&-\sum_{j=1,2}\frac{\alpha_{j}}{2}a_{rj}^{\dagger}a_{rj}+\tilde{H}_{QRj}\label{eq:logic_static},\\
\tilde{H}_{QQ}=&\frac{W}{2}\left(\ket{ee}\bra{gf}e^{2\pi i\nu_r t}+\ket{ee}\bra{fg}e^{2\pi i\nu_r t}\right.\nonumber\\
&\left.+\ket{ee}\bra{gg}e^{2\pi i\nu_b t}+\ket{ee}\bra{ff}e^{2\pi i\nu_b t}\right)\otimes I_{4}
 \nonumber\\
 &+h.c.\nonumber,\\
\tilde{H}_{QR1}=&\frac{\Omega}{2}\left(\ket{eg}\bra{fg}+\ket{ef}\bra{ff}\right)\otimes\ket{0}\bra{1}\otimes I_{2}+h.c. \nonumber,\\
\tilde{H}_{QR2}=&\frac{\Omega}{2}\left(\ket{ge}\bra{gf}+\ket{fe}\bra{ff
}\right)\otimes I_{2}\otimes\ket{0}\bra{1}+h.c. .\nonumber
\end{align}

Finally, we perform another rotating frame transformation so that the detuned QQ sidebands become time-independent, which leads to
\begin{align}
\tilde{H}_{\rm rot}=&-\frac{\alpha_{1}}{2}(P_{eg}+P_{ef})-\frac{\alpha_{2}}{2}(P_{ge}+P_{fe}) \nonumber\\
&-\nu_{r}(P_{gf}+P_{fg}+P_{ge}+P_{eg}) \nonumber\\
&-\nu_{b}(P_{gg}+P_{ff}+P_{ef}+P_{fe})\nonumber\\
&+{\tilde{H}'}_{QQ} \nonumber\\
&-\sum_{j=1,2}\frac{\alpha_{j}}{2}a_{rj}^{\dagger}a_{rj}+\tilde{H}_{QRj}\label{eq:fully_rotated1},\\
\tilde{H}'_{QQ}=&\frac{W}{2}\left(\ket{ee}\bra{gf}+\ket{ee}\bra{fg}\right.\nonumber\\
&\left.+\ket{ee}\bra{gg}+\ket{ee}\bra{ff}+h.c.\right)\otimes I_{4}.\nonumber
\end{align}
Here $I_{n}$ is the $n\times n$ identity matrix and $P_{ab}=\ket{ab}\bra{ab}\otimes I_{4}$. Rotating Wave Approximation (RWA) is applied in the last two transformations. In the final frame, $\{\ket{L_0},\ket{L_1}\}$ have different energies $\{-\nu_r,-\nu_b\}$, and the superposition states become time-dependent.

\section{Circuit Hamiltonian and sideband strength} \label{app:sideband rate}
\begin{figure}[t]
\centering\includegraphics[width=1.0\columnwidth]{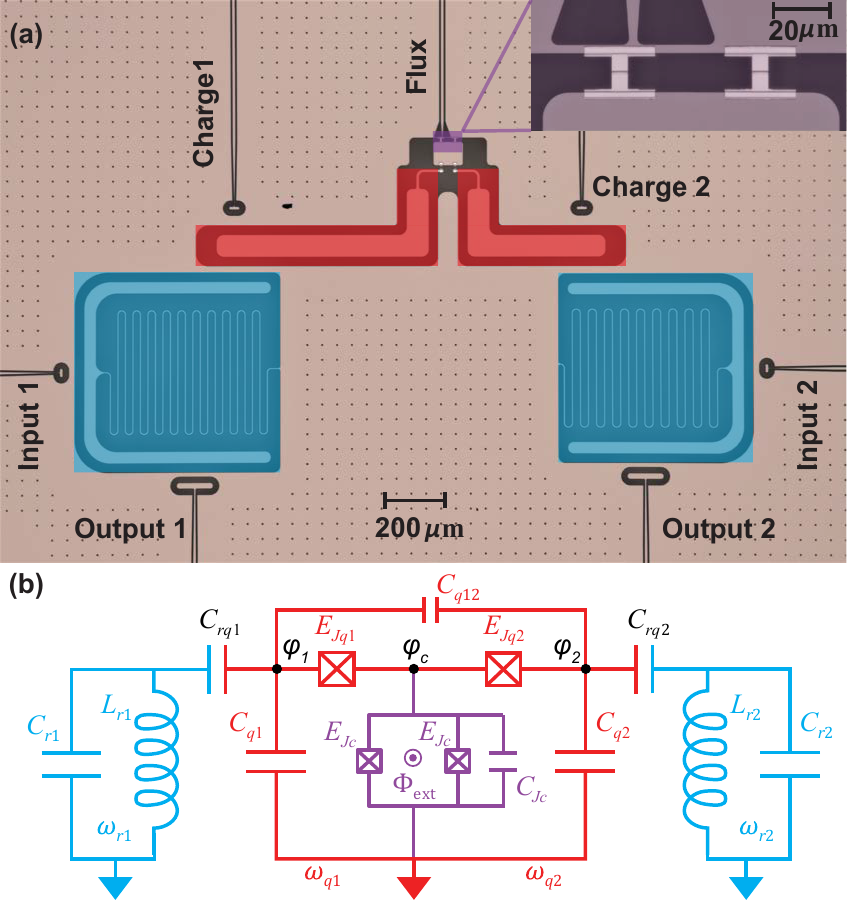}
\caption{The device. (a) False-colored optical image. Two transmons (red) are inductively connected through a SQUID loop (purple, inset shows zoomed-in image). An on-chip flux line is coupled to the SQUID for activating QQ sidebands through parametric RF flux modulation at the proper DC flux position. Each transmon is capacitively coupled to the readout resonator (blue). Single transmon pulses are sent through the resonator input lines. QR sidebands are applied through corresponding charge lines. (b) Circuit schematic diagram.}
\label{fig:Device}
\end{figure}
Fig.~\ref{fig:Device}a is our device's false color optical picture. We first consider the Hamiltonian of the two transmons: 
\begin{subequations}
\begin{align}
    H_{Q} = & \overrightarrow{n}^{\intercal}C_{L}^{-1}\overrightarrow{n} -E_{j1}\cos{\left(\varphi_{c}-\varphi_{1}\right)}-E_{j2}\cos{\left(\varphi_{2}-\varphi_{c}\right)} \nonumber\\ 
    & -E_{jc}\cos{\left(\pi\frac{\Phi_{\rm ext}}{\Phi_0}\right)}\cos{\left(\varphi_{c}\right)} \label{eq:H1}, \\
    C_{L} = &
    \begin{bmatrix}
               C_{q1}+C_{q12} & -C_{q12} & 0 \\
               -C_{q12} & C_{q2}+C_{q12} & 0 \\
               0 & 0 & C_{q1}+C_{q2}+C_{qc}
    \end{bmatrix} \label{eq:H3}, \\
    \overrightarrow{n}^{\intercal} = & \left(n_{1}, n_{2}, n_{c}\right), \; \left[n_{j}, \varphi_{j}\right] = -i.
\end{align}
\end{subequations}    
Here $\overrightarrow{n}$ and $\overrightarrow{\varphi}$ are the charge and phase variables and can be found through the Legendre transformation. Table.~\ref{table:qutrit_params} includes all coefficients used in the quantization. Then we extract the linear part of $H_{Q}$ to obtain
\begin{equation}
\begin{gathered}[b]
    H_{0} = \overrightarrow{n}^{\intercal}C_{L}^{-1}\overrightarrow{n}+\frac{E_{j1}}{2}\left(\varphi_{c}-\varphi_{1}\right)^{2}+\frac{E_{j2}}{2}\left(\varphi_{2}-\varphi_{c}\right)^{2} \\
    +\frac{E_{jc}}{2}\cos{\left(\pi\frac{\Phi_{\rm ext}}{\Phi_0}\right)}\varphi_{c}^{2} \label{eq:H2}.
\end{gathered}   
\end{equation}
Next, we rewrite the charge and phase variables in the dressed basis with the unitary transformation matrix $U$ such that $H_{0}$ is simultaneously diagonalized to find out the normal modes,
\begin{subequations}
\begin{align}
     &H_{0} = \sum_{j=1,2,c}\left(\tilde{C}_{j}\tilde{n}_{j}^{2}+\tilde{D}_{j}\tilde{\varphi_{j}}^{2}\right), \\
     &\overrightarrow{\tilde{n}} = \left(\tilde{n}_{1},\tilde{n}_{2},\tilde{n}_{c}\right)^{\intercal} = U^{-1}\overrightarrow{n}, \\
     &\overrightarrow{\tilde{\varphi}} = \left(\tilde{\varphi}_{1},\tilde{\varphi}_{2},\tilde{\varphi}_{c}\right)^{\intercal} = U^{-1}\overrightarrow{\varphi}, \\
     & U =
    \begin{bmatrix}
               U_{11} & U_{12} & U_{1c} \\
               U_{21} & U_{22} & U_{2c} \\
               U_{c1} & U_{c2} & U_{cc} 
    \end{bmatrix}.
\end{align}
\end{subequations}

In the dressed basis, the nonlinear part is reintroduced in the Hamiltonian to get
\begin{subequations}
\begin{align}
     H_Q &= \sum_{j=1,2,c}\left(\tilde{C}_{j}\tilde{n}_{j}^{2}\right) \nonumber\\
     &-E_{j1}\cos{\left(\sum_{j=1,2,c}\left(U_{cj}\tilde{\varphi}_{j}-U_{1j}\tilde{\varphi}_{j}\right)\right)}  \nonumber\\
     &-E_{j2}\cos{\left(\sum_{j=1,2,c}\left(U_{2j}\tilde{\varphi}_{j}-U_{cj}\tilde{\varphi}_{j}\right)\right)} \nonumber\\ 
     &-E_{jc}\cos{\left(\pi\frac{\Phi_{\rm ext}}{\Phi_0}\right)}\cos{\left(\sum_{j=1,2,c}U_{cj}\tilde{\varphi}_{j}\right)}, \\
     \text{with} \nonumber \\
     \tilde{n}_{j} &=\frac{i}{\sqrt{2}}\sqrt{\frac{\tilde{D}_{j}}{\tilde{C}_{j}}}\left(a_{qj}^{\dagger}-a_{qj}\right), \\
     \tilde{\varphi}_{j} &=\frac{1}{\sqrt{2}}\sqrt{\frac{\tilde{C}_j}{\tilde{D}_j}}\left(a_{qj}^{\dagger}+a_{qj}\right).
\end{align}
\end{subequations}

\begin{figure}[t]
    \centering
    \includegraphics{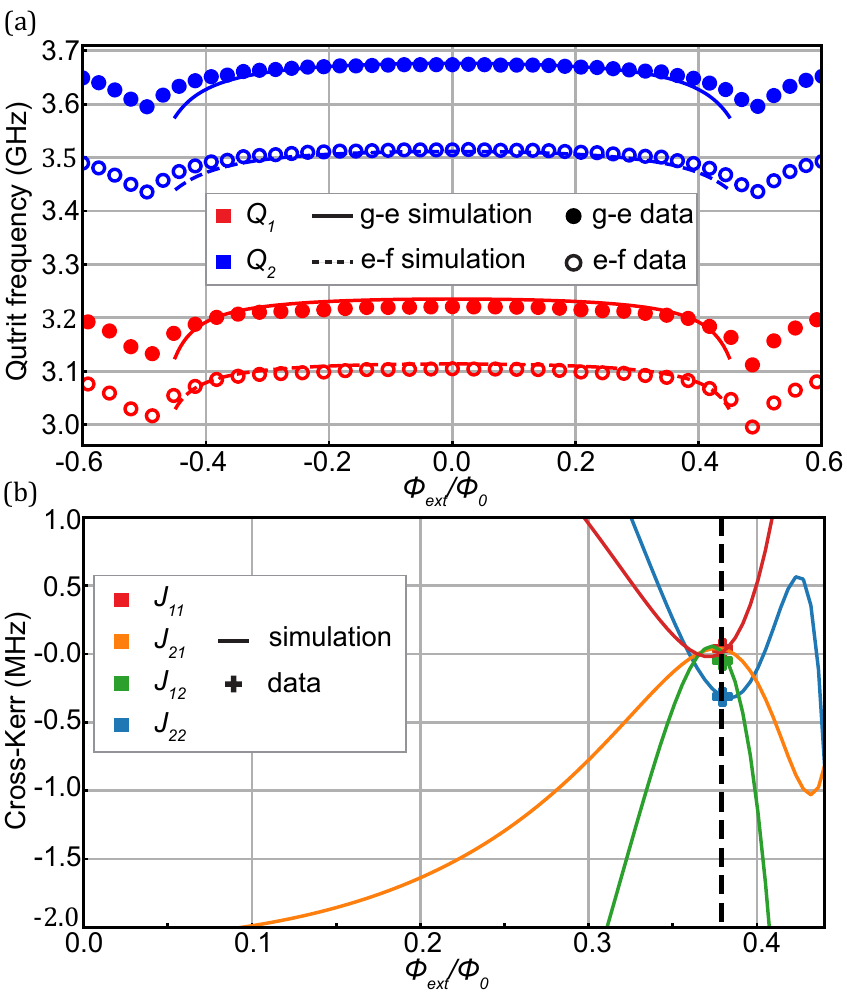}
    \caption{Circuit quantization results of $H_Q$. Comparison of (a) transmon frequencies and (b) cross-Kerr couplings between simulation and experiment. Q1 (red) and Q2's (blue) $\ket{g}\leftrightarrow\ket{e}$ and $\ket{e}\leftrightarrow\ket{f}$ frequencies from numerical calculation and experiment are plotted as a function of $\Phi_{\rm ext}$. Four inter-qutrit cross-Kerr coupling strengths, $J_{11}$, $J_{21}$, $J_{12}$ and $J_{22}$ are calculated, and experiment data are marked out on the Star code operating point (dashed line).}
    \label{fig:quantization}
\end{figure}
\begin{figure}[t]
    \centering
    \includegraphics{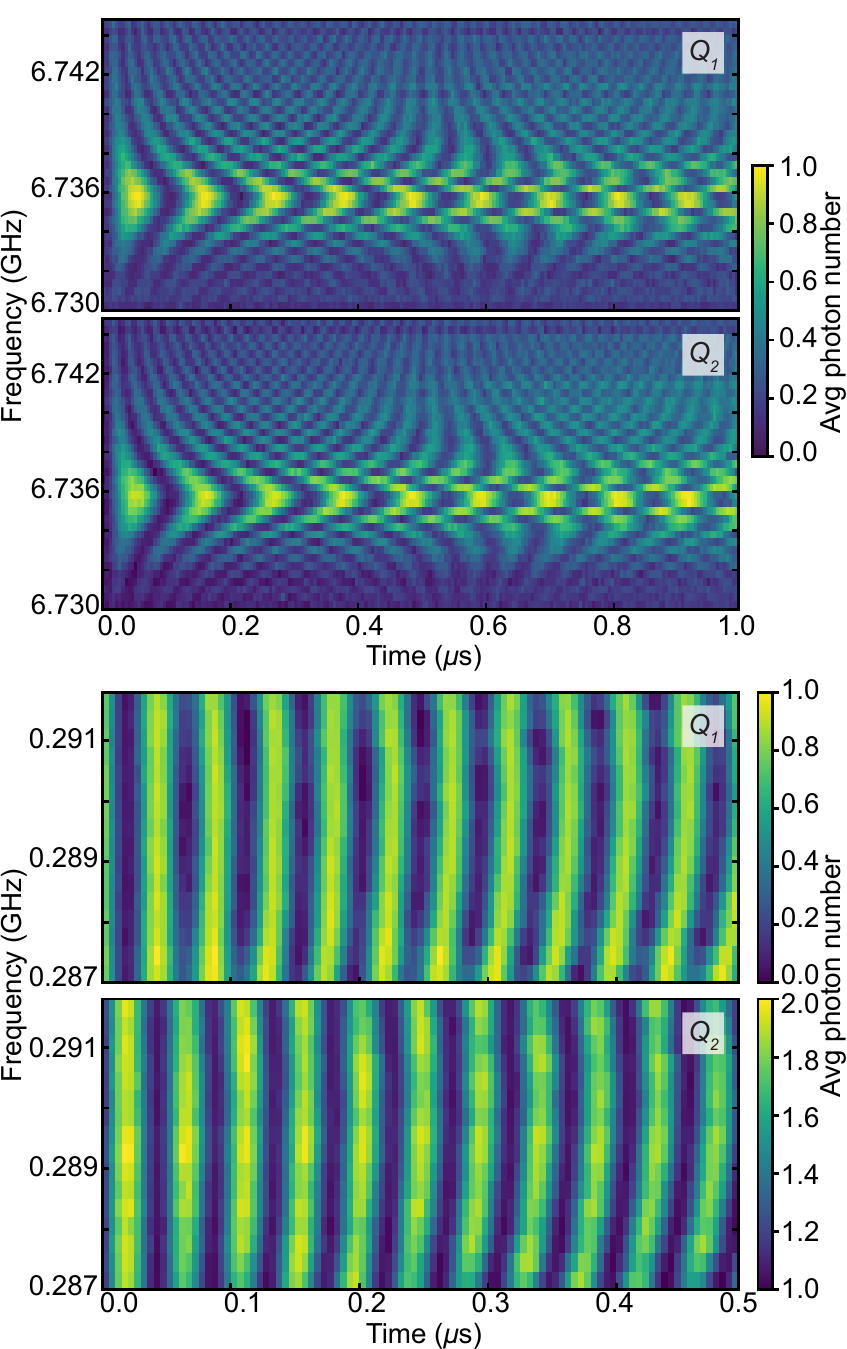}
    \caption{Chevron plots for fast QQ sidebands. The state of both transmons are simultaneously read out and shown as photon numbers. Top two figures show $\SI{9}{\mega\hertz}$ $\ket{gg}\leftrightarrow\ket{ee}$, and bottom two demonstrate $\SI{21}{\mega\hertz}$ $\ket{ee}\leftrightarrow\ket{gf}$ oscillations.}
    \label{fig:supp_readout1}
\end{figure}
We use the scQubits package~\cite{scqubits2022} to quantize the Hamiltonian. The numeric and experiment comparison are shown in Fig.~\ref{fig:quantization}. When $\Phi_{\rm ext}$ is biased close to $\Phi_{0}/2$, deviation appeared in numerics. This comes from the asymmetry of SQUID junctions' resistances and parasitic SQUID loop inductance and is away from our bias point. Around the DC flux position where the Star code protocol is implemented (marked as Fig.~\ref{fig:quantization}b dash line), we have a decent explanation of both transmons' frequencies and cross-Kerr couplings. 

The QQ sidebands are realized through parametric RF flux modulation of the coupler. To understand the sideband rate, we follow the previous paper\cite{Yao2017} and apply an adiabatic approximation to the Hamiltonian: The coupler mode frequency remains high ($>15$ GHz) above transmons' frequencies ($<4$ GHz) in the system, therefore the coupler can be assumed static at the ground state. The non-dynamical potential of the coupler mode is removed by minimizing the Hamiltonian. Transmons are treated as duffing oscillators when calculating the effective sideband rate. Keeping up to 2nd order expansions, the Hamiltonian $H{\rm ad}$ under adiabatic approximation is
\begin{subequations}
\begin{align}
     H_{\rm ad} & = \omega_{q1}a_{q1}^{\dagger}a_{q1}+\omega_{q2}a_{q2}^{\dagger}a_{q2}\nonumber\\
     & +\frac{\alpha_{1}}{2}a_{q1}^{\dagger}a_{q1}^{\dagger}a_{q1}a_{q1}+\frac{\alpha_{2}}{2}a_{q2}^{\dagger}a_{q2}^{\dagger}a_{q2}a_{q2} \nonumber\\
     & +g_{1}\left(t\right)\left(a_{q1}^{\dagger}+a_{q1}\right)\left(a_{q2}^{\dagger}+a_{q2}\right) \nonumber\\
     & +g_{2}\left(-a_{q1}^{\dagger}+a_{q1}\right)\left(-a_{q2}^{\dagger}+a_{q2}\right), \\
     g_{1}\left(t\right) = & \frac{\sqrt{E_{j1}E_{j2}}}{2E_{jc}\cos{\left(\pi\frac{\Phi_{\rm ext}\left(t\right)}{\Phi_0}\right)}}\sqrt{\omega_{q1}\omega_{q2}}, \label{eq:quantization} \\
     g_{2} = & \frac{\sqrt{C_{q1}C_{q2}}}{2C_{q12}}\sqrt{\omega_{q1}\omega_{q2}}. 
\end{align}
\end{subequations}
Here $g_{1}\left(t\right)$ and $g_{2}$ are flux-tunable inductive coupling strength and constant capacitive coupling strength. Plugging in the RF flux modulation $\frac{\pi\Phi_{\rm ext}\left(t\right)}{\Phi_0}=\Phi_{\rm DC}+\epsilon\cos\left(\omega_{d}t\right)$ into Eq.~\ref{eq:quantization} and assuming $\epsilon\ll\Phi_{\rm DC}$, we obtain 
\begin{align}
     g_{1}\left(t\right) = & \frac{\sqrt{E_{j1}E_{j2}}}{2E_{jc}}\sqrt{\omega_{q1}\omega_{q2}}\frac{1}{\cos\left(\Phi_{\rm DC}+\epsilon\cos\left(\omega_{d}t\right)\right)} \nonumber \\
     = & \frac{\sqrt{E_{j1}E_{j2}}}{2E_{jc}}\sqrt{\omega_{q1}\omega_{q2}}\frac{\left(1+\epsilon\sin\left(\omega_{d}t\right)\tan\left(\Phi_{\rm DC}\right)\right)}{\cos\left(\Phi_{\rm DC}\right)}.
     \label{eq:anl rate}
\end{align}
Therefore the QQ sideband rate becomes (suppose $\ket{\psi_1}$ and $\ket{\psi_2}$ are states connected by the sideband)
\begin{align}
&\frac{\sqrt{E_{j1}E_{j2}}}{2E_{jc}}\sqrt{\omega_{q1}\omega_{q2}}\frac{\epsilon\tan\left(\Phi_{\rm DC}\right)}{\cos\left(\Phi_{\rm DC}\right)}A_{12}, \\
\text{with} \nonumber \\
&A_{12} = \bra{\psi_1}\left(a_{q1}^{\dagger}+a_{q1}\right)\left(a_{q2}^{\dagger}+a_{q2}\right)\ket{\psi_2},\nonumber
\end{align}
and is proportional to the flux modulation rate. $A_{12}$ is the state-dependent bosonic enhancement coefficient. Higher order corrections can be calculated using time-dependent Schrieffer–Wolff transformation\cite{IBM2018}, and for our inductive coupler, both QQ blue and red sideband will have a similar rate under the same $\epsilon$.

Each transmon is capacitively coupled to the readout resonator, and the second-order QR error-correcting sidebands are generated through the charge drive at half the transition energy $\omega_{dqrj}=(\omega_{qj}+\omega_{rj}+\alpha_{qj})/2$ with drive amplitude $\epsilon_{qj}$,
\begin{align}
     H_{QRj} = & \omega_{qj}a_{qj}^{\dagger}a_{qj}+\frac{\alpha_{j}}{2} \left(a_{qj}^{\dagger}\right)^2 \left(a_{qj}\right)^2+\omega_{rj}a_{rj}^{\dagger}a_{rj} \nonumber\\
     & +g_{qrj}\left(-a_{qj}^{\dagger}+a_{qj}\right)\left(-a_{rj}^{\dagger}+a_{rj}\right) \nonumber \\ 
     & + \epsilon_{qj}\left(a_{qj}^{\dagger}e^{-i\omega_{dqrj}t}+h.c.\right) .
\end{align}

The effective QR sideband rate is $\Omega_{j}={16g_{qrj}^{3}\epsilon_{qj}^{2}}/{\left(\omega_{qj}-\omega_{rj}\right)^4}$~\cite{wallraff2007sidband}. Under the same Purcell limit $\left(\sim\frac{g_{qrj}^2\kappa_j}{\left(\omega_{qj}-\omega_{rj}\right)^{2}}\right)$ from the resonator, smaller QR frequency difference allows higher QR sideband rate for the same drive amplitude. In our experiment, the Purcell limit from the resonator is larger than $\SI{200}{\micro\second}$, and is not limiting the physical coherence time.

\section{Transmon $ZZ$ induced dephasing}
\label{section:ZZ}
 Realizing AQEC requires error transparency to single photon loss error. This makes $ZZ$ coupling an extra logical dephasing channel as it does not commute with $\tilde{H}_{star}$. In a two qutrit system, there are in total 7 different $ZZ$ frequency shifts coming from 4 cross-Kerr coupling strengths $J_{11}, J_{21}, J_{12}, J_{22}$. However, not all $ZZ$ couplings are detrimental to the Star code. The error transparency requires no phase accumulation between logical states during the error correction process. This is equivalent to having the same energy for the photon lost from one transmon, independent of the state of the partner transmon,
\begin{equation}
\begin{split}
     E_{ff}-E_{ef} &= E_{fg}-E_{eg}, \\
     E_{ff}-E_{fe} &= E_{gf}-E_{ge}.
     \label{eq:ZZ_requirement}
\end{split}
\end{equation}
Here $E_{jk}$ refers to the energy of the state $\ket{jk}$. Equations~\ref{eq:ZZ_requirement} are equivalent to $ZZ_{ff2}=ZZ_{ff1}=0$ (see Table.~\ref{table:frequency}). When this is not the case, a random phase difference will accumulate between logical states after the error correction, introducing dephasing to the logical superposition states. The other $ZZ$s are naturally error transparent in the logical manifold since $\ket{ee}$ is dark and will not affect the logical states' coherence. To suppress the $ZZ$-induced logical dephasing, we can increase the QR sideband rate $\Omega_{j}$, shortening the $\ket{e}$ population time in both qutrits and reducing the accumulated random logical phase. To eliminate this dephasing channel, we need a coupler that has zero $ZZ_{ff1}$ and $ZZ_{ff2}$ when all external sidebands are turned on. This can be potentially realized in our current coupler with dispersive shift engineering. We consider a two-transmon system with static interaction that produces a set of dispersive shifts for cancellation. The base Hamiltonian $H_{\rm base}$ is
\begin{equation}
\begin{split}
     H_{\rm base} = & \sum_{n}\left(\epsilon_{1,n}\ket{n_1}\bra{n_1}+\epsilon_{2,n}\ket{n_2}\bra{n_2}\right) \\
     & + \sum_{nm}\Delta_{nm}\ket{n_1m_2}\bra{n_2m_1},
     \label{eq:base_H}
\end{split}
\end{equation}
where $\epsilon_{1/2,n}$ are energies for single transmon levels, $\Delta_{nm}$ is the static energy shift to the state when transmon $1$ has $n$ photons and transmon $2$ has $m$ photons. For the ground state $\epsilon_{1/2,0}$ and $\Delta_{00}$ are set to 0. We add to $H_{\rm base}$ a QQ red sideband through the coupler,
\begin{equation}
\begin{split}
    H_{D}=2g\sin{\left(2\pi\nu t\right)}\left(a_{q1}^{\dag}a_{q2}+a_{q1}a_{q2}^{\dag}\right).
     \label{eq:Hd}
\end{split}
\end{equation}
We assume the frequency detuning $\nu$ is far off-resonant so that $\nu\gg g$, and this will introduce an energy shift $D_{jk}$ to all levels:
\begin{equation}
\begin{split}
     D_{jk}^{(R)} = & \frac{g^{2}j(k+1)}{E_{j,k}-E_{j-1,k+1}-\nu}+\frac{g^{2}j(k+1)}{E_{j,k}-E_{j-1,k+1}+\nu} \\
     & + \frac{g^{2}(j+1)k}{E_{j,k}-E_{j+1,k-1}-\nu}+\frac{g^{2}(j+1)k}{E_{j,k}-E_{j+1,k-1}+\nu}.
     \label{eq:ZZ_engineering_red}
\end{split}
\end{equation}
For a detuned QQ blue sideband drive, one can find a similar expression for the energy shift,
\begin{equation}
\begin{split}
     D_{jk}^{(B)} = & \frac{g^{2}(j+1)(k+1)}{E_{j,k}-E_{j+1,k+1}-\nu}+\frac{g^{2}(j+1)(k+1)}{E_{j,k}-E_{j+1,k+1}+\nu} \\
     & + \frac{g^{2}jk}{E_{j,k}-E_{j-1,k-1}-\nu}+\frac{g^{2}jk}{E_{j,k}-E_{j-1,k-1}+\nu}.
     \label{eq:ZZ_engineering_blue}
\end{split}
\end{equation}
\begin{figure}[t]
    \centering
    \includegraphics{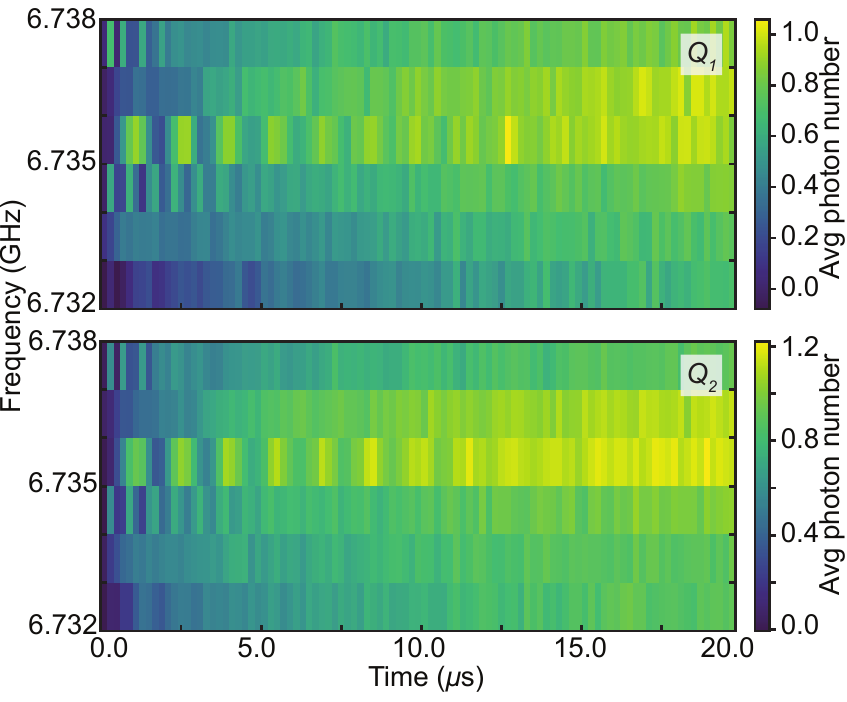}
    \caption{Readout saturation feature for fast (rate $\SI{9}{\mega\hertz}$) QQ blue sideband  $\ket{gg}\leftrightarrow\ket{ee}$ in the long time scale. Top and bottom panel show readout signals from the first and second resonators.}
    \label{fig:supp_readout2}
\end{figure}
When multiple external QQ sidebands are applied, the total dispersive shift to each energy level is given by $D_{jk} = \sum D_{jk}^{(R)} + \sum D_{jk}^{(B)}$, where the sum is over all external QQ sidebands. Specifically for the Star code scheme, we can modulate two extra QQ sidebands near $\ket{ef}\leftrightarrow\ket{gh}$ and $\ket{fe}\leftrightarrow\ket{hg}$. This can introduce either positive or negative $D_{12}$ and $D_{21}$ to the system, depending on the choices of frequency detuning. The required $g$ for complete $ZZ$ cancellation can be less than $\SI{5}{\mega\hertz}$. Therefore, it is theoretically possible to cancel $ZZ_{ff1}$ and $ZZ_{ff2}$ simultaneously under such two extra drives. We did not turn on the cancellation sidebands in the experiment, because our readout suffers from frequency shift under strong flux modulation amplitude as discussed in Appendix~\ref{app:readout problem}.
\section{Readout saturation and flux line optimization}\label{app:readout problem}
\begin{figure}[t]
    \centering
    \includegraphics{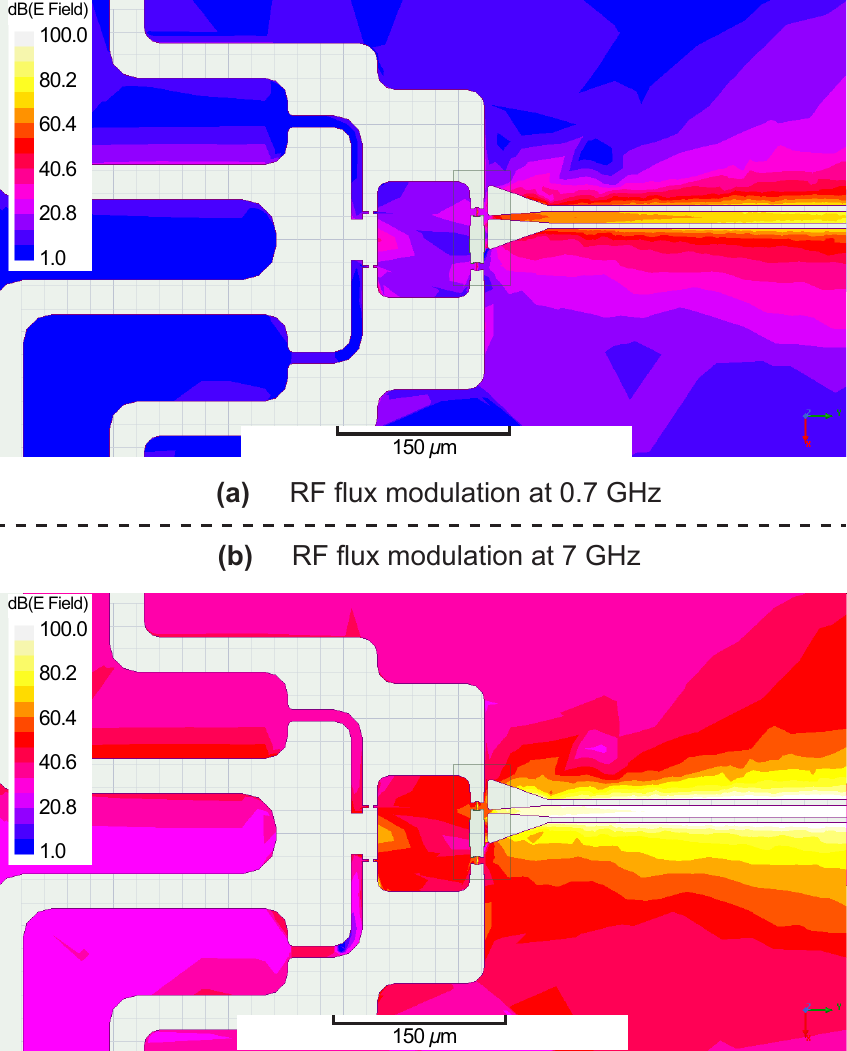}
    \caption{Distribution of electrical field's amplitude (in log scale) when RF flux drive is modulated at (a) 700 MHz and (b) 7 GHz obtained from HFSS simulation.}
    \label{fig:hfss_sim}
\end{figure}
In our device, we use the on-chip flux line to generate various two-qubit interactions through parametric modulation. A typical parametric coupler design includes two qubits capacitively or inductively coupled through a tunable coupler. Modulation of the coupler frequency and the qubit-coupler coupling strength contribute to the two-qubit interaction strength. For a capacitively coupled system~\cite{IBM2018, 2020chuji, IBMcalib2020, 2021rigetti}, coupler frequency modulation contributes dominantly to the QQ sideband rate, and time-dependent Schrieffer-Wolff transformation (SWT) proves that~\cite{IBM2018} the ratio of interaction strengths between bswap and iswap is $\frac{\omega_{1}-\omega_{2}}{\omega_{1}+\omega_{2}}$ ($\omega_{j}$ is $Q_j$'s frequency). Therefore, a capacitive coupler provides a slower bswap than the iswap. In contrast, an inductively coupled system~\cite{Yao2017, Brown2022} modulates the coupling strength between the qubit and coupler more effectively, and both iswap and bswap will have the same zeroth-order terms in the SWT expansion, thus theoretically sharing the same rate under same modulation amplitude. Previous experiments achieved fast iswap interactions, but a similar bswap rate has not been demonstrated in either type of parametric coupler yet. We experimentally realize a comparable maximum rates of $\SI{9}{\mega\hertz}$ bswap and $\SI{21}{\mega\hertz}$ iswap (shown in Fig.~\ref{fig:supp_readout1}).
\begin{figure}[t]
    \centering
    \includegraphics{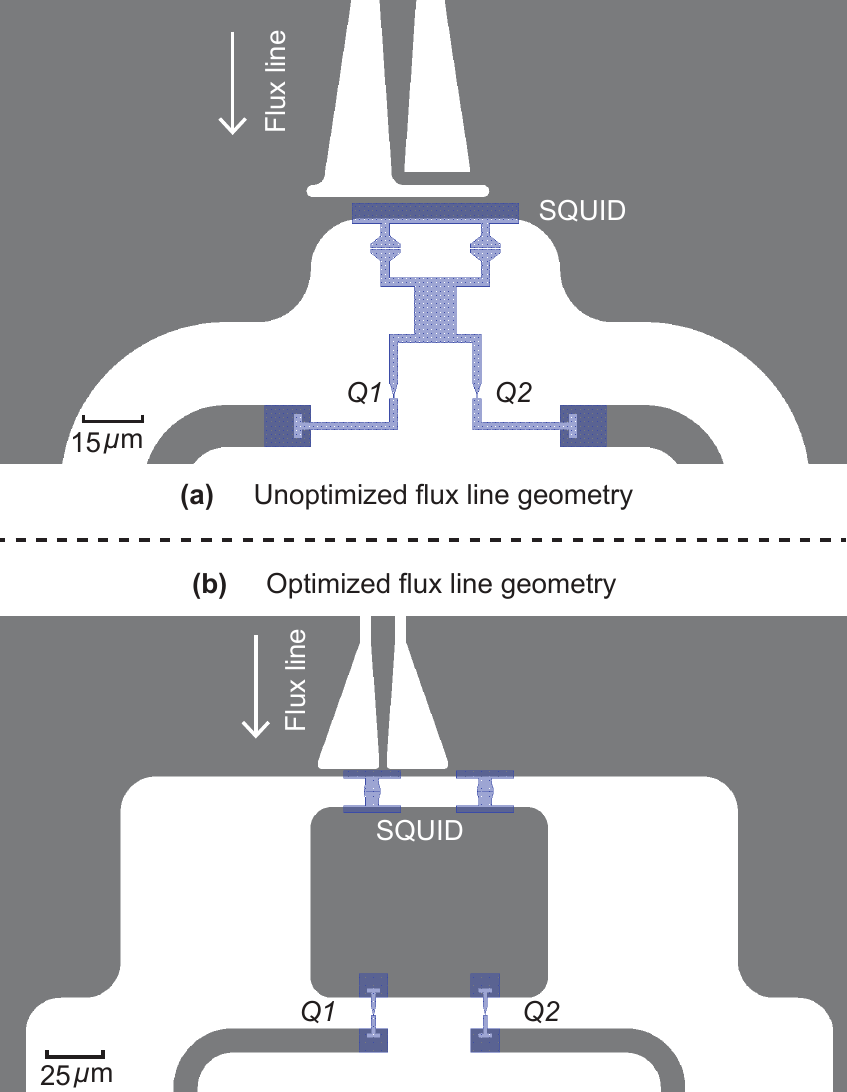}
    \caption{SQUID design. (a) Old design without optimization. (b) Current design by maximizing the mutual inductance between the flux line and SQUID. The Ta and Al areas are colored as grey and blue separately, and the bare sapphire is colored white.}
    \label{fig:flux_geometry}
\end{figure}

In the experiment, we notice that turning a strong bswap on will shift both resonators' frequencies after a long time, resulting in the `saturation' feature (Fig.~\ref{fig:supp_readout2}).
Such a readout frequency shift is both sideband strength and duration dependent, and the shift persists for a noticeable period after all sidebands are turned off.  Distinguishing transmons' states through readout becomes difficult when this happens. A readout is possible when the shift is reversed after waiting for a sufficiently long period but degrades readout fidelity due to transmons' relaxation. While case-dependent dynamic demarcation can distinguish states, this method becomes complex and inaccurate. In our experiment, we decided to lower the RF modulation amplitude and minimize the saturation region by optimizing the flux line geometry.

One source for the readout saturation at the bswap drive frequency is the flux line's stray charge coupling to the SQUID~\cite{Brown2022}. The on-chip flux line can be considered an antenna. The coupler is located in the near-field region, and the electrical field amplitude is proportional to the flux modulation frequency. Since bswap's drive frequencies are normally a magnitude higher than that of the iswap operations, a much stronger stray-charge drive is observed when the bswap drive is on.
\begin{figure}[t]
    \centering
    \includegraphics{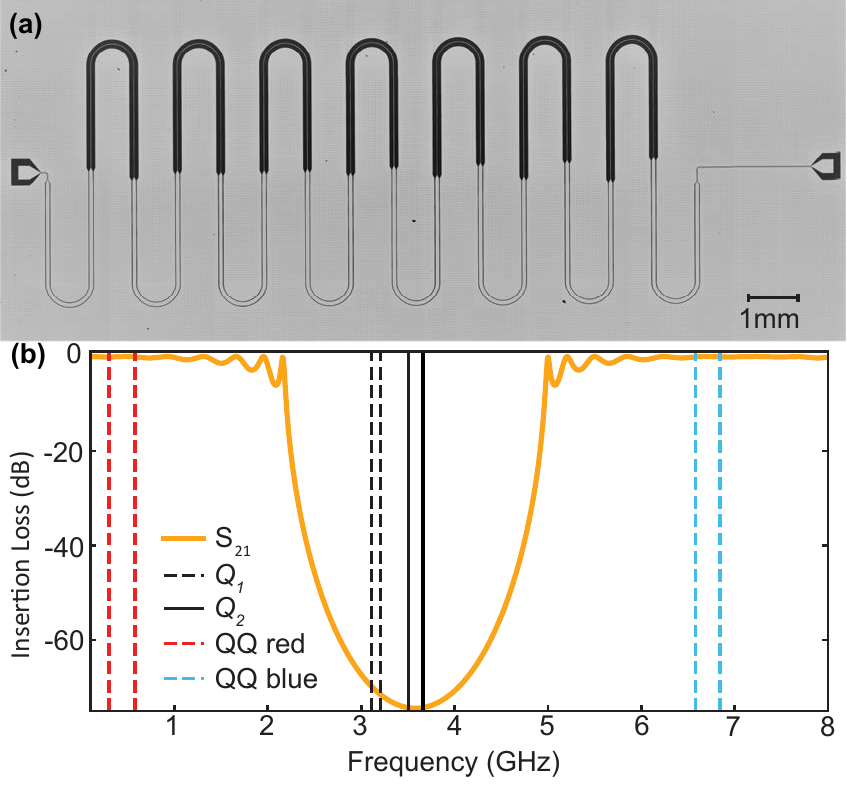}
    \caption{The stepped impedance Purcell filter. (a) Optical image of our SIPF chip made using Ta on sapphire. (b) Calculated SIPF insertion loss. Transmon transitions, QQ red and blue sideband frequencies are marked in the plot.}
    \label{fig:flux_filter}
\end{figure}

We verify this fact using ANSYS HFSS simulation (see Fig.~\ref{fig:hfss_sim}), where the electrical field amplitude is observed to increase over an order of magnitude when the modulation frequency is increased by a factor of 10. The stray charge drive limits the maximum power we can use for the flux modulation, and we focus on geometrical optimization to improve the flux-to-charge drive ratio. In order to do so, we maximize the mutual inductance between the SQUID and the flux line by increasing the SQUID loop size and bringing the flux line closer to the loop. The loop size in our experiment is limited by the SQUID's hysteresis~\cite{sQUID_1976} set by the ratio $\frac{L_{\rm loop}}{L_{jc}}$, where $L_{jc}$ is the inductance of each junction in the SQUID (assumed identical) and $L_{\rm loop}$ is the SQUID loop inductance. When $\frac{L_{\rm loop}}{L_{jc}}>1$, transmon frequencies become hysteric as a function of $\Phi_{\rm DC}$, and the region grows with the ratio. Dissipation appears when modulating within the hysteric region and should be avoided in our experiment. This property sets an upper bound for SQUID loop length. 

We use HFSS simulation to calculate flux threaded by the SQUID loop and vary the geometry to maximize. Assuming the electrical field is geometry insensitive around the SQUID, maximizing SQUID flux increases the ratio between the mutual inductive coupling and stray capacitive coupling strength of the flux line. The original and optimized designs are shown in Fig.~\ref{fig:flux_geometry}. The simulation suggests a factor of $3.5$ improvement in the ratio.

Being strongly coupled to the SQUID, the flux line is also a channel for transmons' relaxation. In order to improve Purcell protection, we design a Stepped-Impedance band-stop Purcell Filter (SIPF) as shown in Fig.~\ref{fig:flux_filter}, which strongly blocks transmon frequencies while allowing the QQ red and blue sideband drives to pass (see Fig.~\ref{fig:measurement} for the full measurement setup). 

\section{QQ and QR sideband rates}
\label{app:sidebandexp}
\begin{figure}[t]
    \centering
    \includegraphics{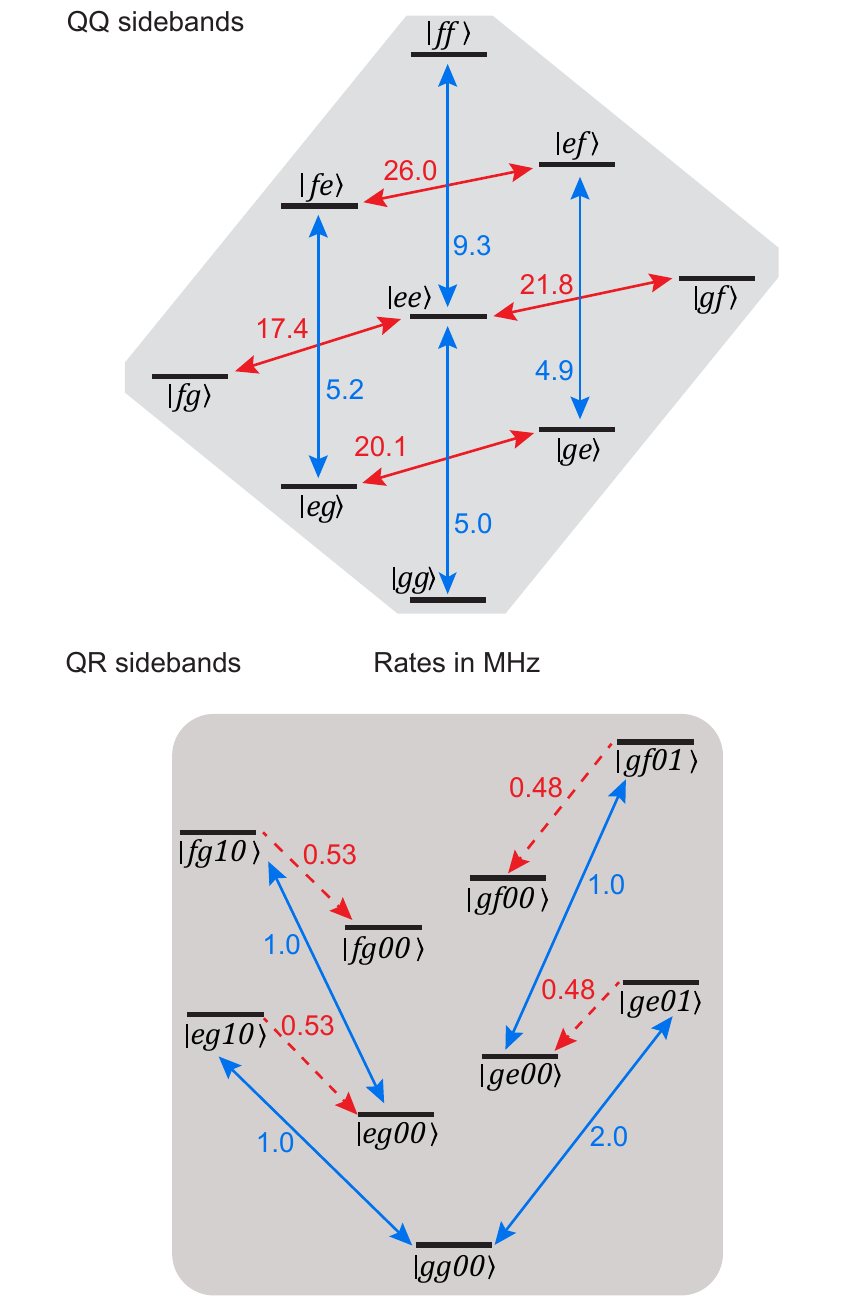}
    \caption{QQ (top panel) and QR (bottom panel) sidebands with rates achieved in the experiment without readout saturation.}
    \label{fig:sideband_rate}
\end{figure}

The ability to sustain fast QQ sidebands is crucial to the realization of the Star code and we optimize the geometry of the device as discussed in the previous section. In Fig.~\ref{fig:sideband_rate}(a), we show all two-photon QQ sidebands achieved in the experiment. Fig.~\ref{fig:sideband_rate}(b) shows all experimentally achieved QR sidebands and resonator decay rates. The strong and rich two-photon processes with this design also show the potential of realizing high-fidelity two-qutrit gates~\cite{2Qutrit2022}.

\section{Full Star code calibration process}
\label{app:calibration}
\begin{figure}[t]
    \centering
    \includegraphics[width=\columnwidth]{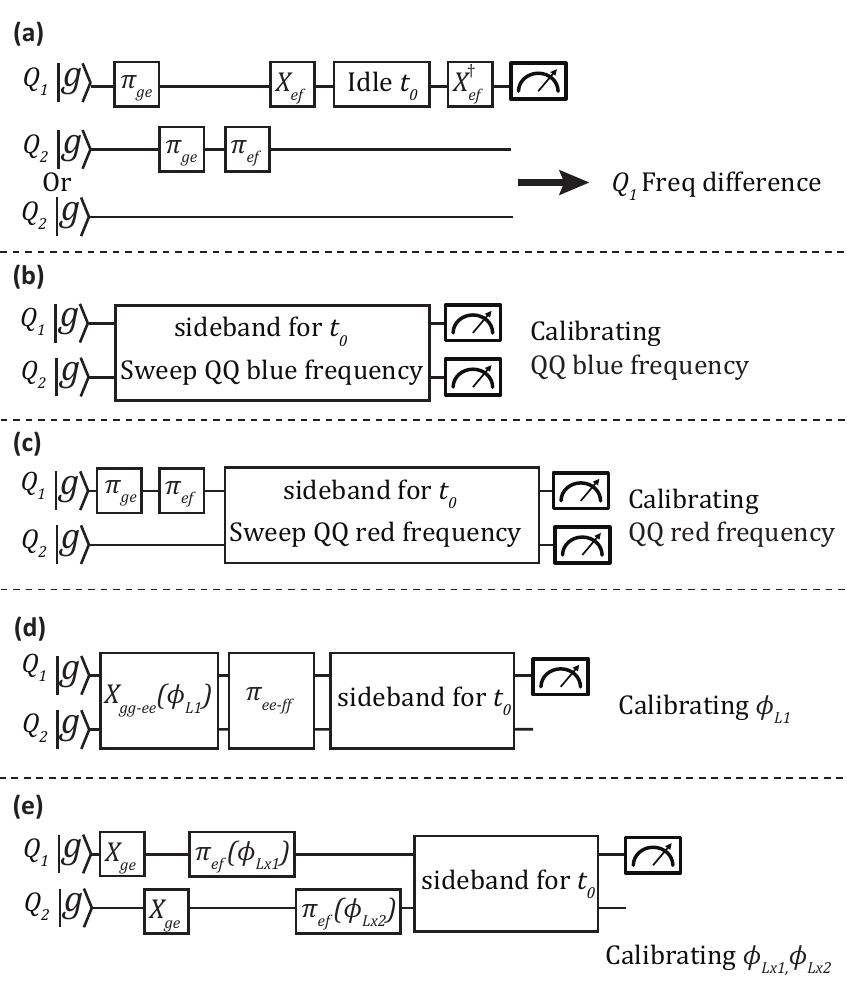}
    \caption{Gate circuits for the Star code calibration. (a) Pulse sequence for calibration of $ZZ_{ff1}$. Ramsey-like protocol is used for the partner transmon being in $\ket{g}$ or $\ket{f}$. The other $ZZ$s are calibrated similarly. (b) QQ blue and (c) QQ red sideband frequency calibration sequence. QQ sideband pair frequencies are iteratively swept and updated based on the time-domain pattern. (d) $\ket{L_1}$ preparation phase calibration sequence. The phase point that has minimum $\ket{e}$ population in the sweep is chosen as the calibrated $\phi_{L1}$. The preparation phase for $\ket{L_0}$ is calibrated similarly. (e) $\ket{L_x}$ preparation phase calibration protocol.}
    \centering
    \label{fig:calibration_circuit}
\end{figure}
To implement the Star code, we need to calibrate the QQ and QR sideband frequencies when all sidebands are simultaneously on. The presence of external sidebands will change both transmons' frequencies through AC-stark shift and rectifying effect (RF modulation under a nonlinear frequency-flux response). In the experiment, we systematically perform the calibration, shown in Fig.~\ref{fig:calibration_circuit}.
\begin{figure}[t]
    \centering
    \includegraphics[width=\columnwidth]{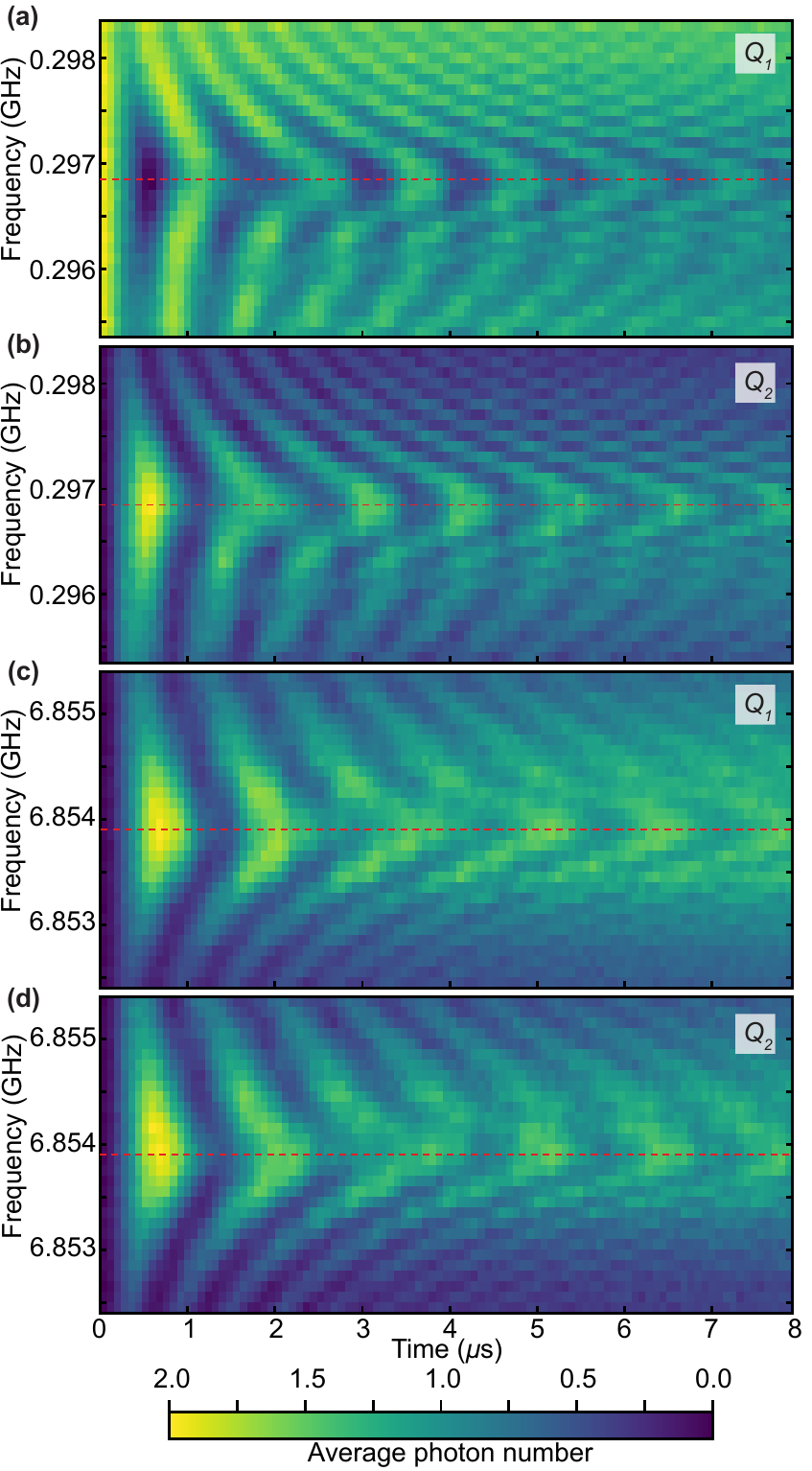}
    \caption{Simultaneous QQ sideband calibration. All six tones used in the Star code are simultaneously turned on during the frequency-time sweep. The red (top two panels) and blue (bottom two panels) QQ sideband pair frequencies are swept individually for calibration in the presence of the other pair. Red dash lines in the plots represent the pairs' center frequency choices. (a) and (b) are $Q_{1}$ and $Q_{2}$'s average photon number when sweeping the red pair with the initial state  $\ket{fg}$. (c) and (d) are $Q_{1}$ and $Q_{2}$'s average photon number when sweeping the blue pair with the initial state $\ket{gg}$. }
    \centering
    \label{fig:sweep_block_calibraion}
\end{figure}
\begin{figure}[t]
    \centering
    \includegraphics[width=\columnwidth]{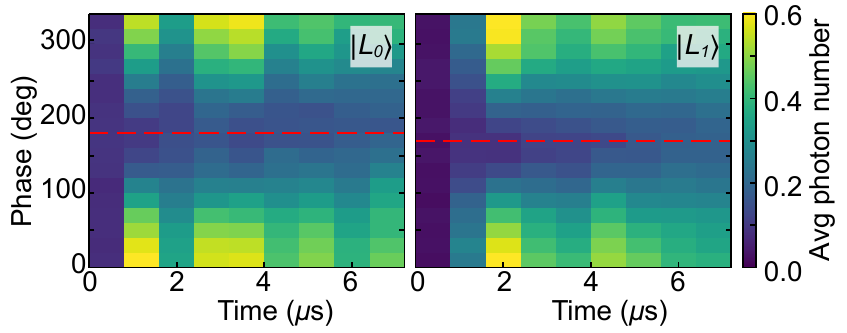}
     \caption{Preparation phase calibration of $\ket{L_0}$ (left) and $\ket{L_1}$ (right) in the error correction experiment. The population of $\ket{e}$ on $Q_{1}$ are measured after turning on all sidebands for up to $\SI{8}{\micro\second}$, and the red dash line marks the calibrated phase position. We choose Q1 due to higher readout fidelity.}
    \centering
    \label{fig:L0/1 phase_calibration}
\end{figure}

\begin{figure}[b]
    \centering   \includegraphics[width=\columnwidth]{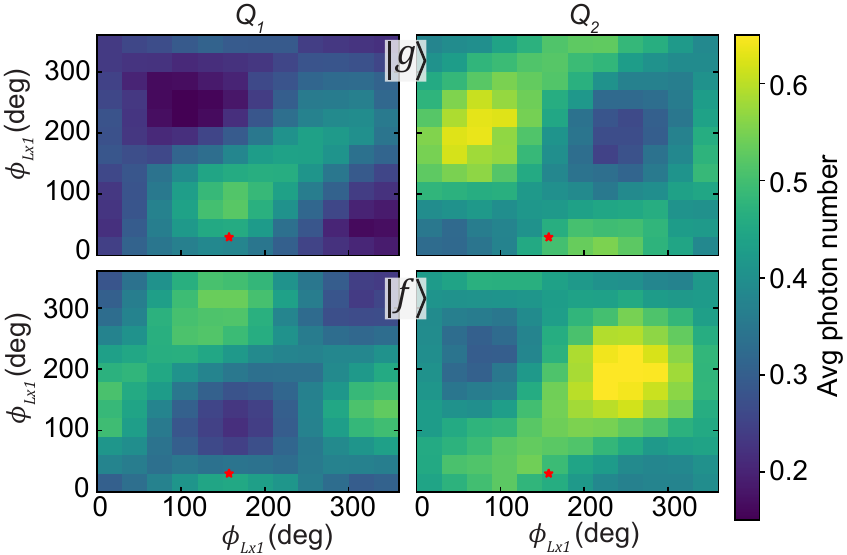}
    \caption{Phase calibration for $\ket{L_x}$ preparation. Population of $\ket{g}$ (top)  and $\ket{f}$ (bottom) on $Q_{1}$ (left) and $Q_{2}$ (right) are measured after $\SI{8}{\micro\second}$, and the red star marks the calibrated phase position.}
    \centering
    \label{fig:Lx phase_calibration}
\end{figure}
In Fig.~\ref{fig:calibration_circuit}a, the static $ZZ$ dispersive shift is characterized by measuring the Ramsey fringe frequency difference depending on the other qubit's state. In steps Fig.~\ref{fig:calibration_circuit}b and Fig.~\ref{fig:calibration_circuit}c, we first turn on all 6 QQ and QR drives at their bare frequencies. All QQ sideband rates are set to $W$ when independently turned on. Two QQ red sidebands and two QQ blue sidebands are pair-swept separately as the `red/blue pair'. The pair width and center are the sidebands' frequency difference and average. In each iteration step, we update sequentially the red and blue pair centers, and the QR sideband frequencies. For each pair, we sweep the center frequency as a function of time with all six sidebands on. We use $\ket{gg}$ (blue pair) and $\ket{fg}$ (red pair) as the initial states. Reading out the average photon number in both transmons, the 2D sweep plots show a fringed chevron pattern (shown in Fig.~\ref{fig:sweep_block_calibraion}). The pattern's center line is the new pair center. The fringe rate represents the actual sideband detunings $\nu_{b/r}$ and rate $W_{b/r}$, and the detunings can be updated by changing pair width at this stage. After extracting both pairs' new centers, the QR sidebands are calibrated with $\ket{eg}$ and $\ket{ge}$ as the initial state when all drives are on. Populating $\ket{f}$ with the $\ket{e0}\leftrightarrow\ket{f1}$ process is most efficient when QR sidebands are on resonance. Because of none zero $ZZ_{ff1}$ and $ZZ_{ff2}$, the QR sidebands cannot be exactly on resonance for both $\ket{L_i}$. In the experiment, we calibrate QR sidebands to be on resonance for the $\ket{L_0}$. For $\ket{L_1}$, the error correction process will be slower but not dephase the state after correction. After a few iterations, we get decent frequency calibrations of all six sidebands.

Logical state preparation includes both charge and flux drives with appropriate relative phases. Fig.~\ref{fig:calibration_circuit}d is to calibrate $\ket{L_0}$ and $\ket{L_1}$'s preparation phase. For the logical state $\ket{L_0}$, we first apply two $\pi_{ge}$ pulses sequentially on $Q_{1}$ and $Q_{2}$ to prepare $\ket{ee}$ through charge lines. Afterwards a $(\pi/2)_{\ket{ee}\leftrightarrow \ket{fg}}$ pulse with a phase offset $\phi_{L0}$ and a $\pi_{\ket{ee}\leftrightarrow\ket{fg}}$ pulse are applied through the flux line. To prepare $\ket{L_1}$, a $(\pi/2)_{\ket{gg}\leftrightarrow \ket{ee}}$ pulse with some phase $\phi_{L1}$ followed by a $\pi_{\ket{ee}\leftrightarrow\ket{ff}}$ are applied through the flux line. These steps generally prepare $(\ket{gf}-e^{i\phi_{L0}}\ket{fg})/\sqrt{2}$ and $(\ket{gg}-e^{i\phi_{L1}}\ket{ff})/\sqrt{2}$. For non-zero $\{\phi_{L0}, \phi_{L1}\}$, $\ket{ee}$ is populated under the action of $\tilde{H}_{\rm static}$, and we use this feature to find the correct preparation phases. We sweep the phase $\phi_{L0(1)}$ in the presence of all six tones and observe $\ket{e}$ population on both qutrits. The correct preparation phases are determined by values that minimize $\ket{e}$ population of both transmons during the first $\SI{8}{\micro\second}$ of error correction, as presented in Fig.~\ref{fig:L0/1 phase_calibration}.

Preparation of $\ket{L_{x}} = (\ket{L_0}+\ket{L_1})/\sqrt{2}$ does not require any sideband pulses as it is a product state $\left(\ket{g}-\ket{f}\right)\left(\ket{g}+\ket{f}\right)/2$. We apply a $(\pi/2)_{ge}$ pulse with a specific phase, followed by a $\pi_{ef}$ pulse on both transmons. These pulses prepare the state $\left(\ket{g}+e^{i\phi_{Lx1}}\ket{f}\right)\left(\ket{g}+e^{i\phi_{Lx2}}\ket{f}\right)/2$, leaving two preparation phases $\phi_{Lx1}$ and $\phi_{Lx2}$ left for calibration. The correct phase combination can be calibrated on the 2D $\phi_{Lx1}$-$\phi_{Lx2}$ phase sweep plot. Correct preparation phases will keep equal populations of $\ket{g}$ and $\ket{f}$ for both transmons at any time after turning on all sidebands. In Fig.~\ref{fig:calibration_circuit}e, both transmons' $\ket{g}$ and $\ket{f}$ populations are measured $\SI{8}{\micro\second}$ after turning sidebands on. Fig.~\ref{fig:Lx phase_calibration} shows the 2D phase sweep plot. This yields four phase coordinates $\{\phi_{Lx1},\phi_{Lx2}\}=\{0,180^\circ\}\otimes\{0,180^\circ\}$, and two of the four correspond to $(\ket{L_0}\pm\ket{L_1})/\sqrt{2}$. We distinguish the logical $\ket{L_x}$ by taking two-qutrit state tomography measurements after turning on the sidebands for $\SI{9}{\micro\second}$ and choose the error-corrected case. The calibration process for the 4 QQ echo case is the same, except the QR sidebands are off.

\section{Two qutrit tomography}
\label{app:tomography}
Following the basis choice in Ref.~\onlinecite{Wallraff2010qutrit_tomo}, we apply 81 post rotations $S_{j}$ from the tomography rotation set $S\otimes S$: $S=\{I, R_{ge}\left(0,\frac{\pi}{2}\right), R_{ge}\left(\frac{\pi}{2},\frac{\pi}{2}\right),R_{ge}\left(0,\pi\right),R_{ef}\left(0,\frac{\pi}{2}\right),$\\
$R_{ef}\left(\frac{\pi}{2},\frac{\pi}{2}\right),R_{ef}\left(0,\frac{\pi}{2}\right)R_{ge}\left(0,\pi\right),R_{ef}\left(\frac{\pi}{2},\frac{\pi}{2}\right)R_{ge}\left(0,\pi\right),$\\
$R_{ef}\left(0,\pi\right)R_{ge}\left(0,\pi\right)\}$. Here $I$ is the identity gate, and the rotations $R_{ge}$ and $R_{ef}$ are defined as follows
\begin{subequations}
\label{eq:single_qutrit}
\begin{align}
    R_{ge}\left(\phi,\theta\right) &= 
    \begin{bmatrix}
               \cos{\frac{\theta}{2}} & -e^{-i\phi}\sin{\frac{\theta}{2}} & 0 \\
           e^{i\phi}\sin{\frac{\theta}{2}} & \cos{\frac{\theta}{2}} & 0 \\
           0 & 0 & 1
    \end{bmatrix},
    \\
    R_{ef}\left(\phi,\theta\right) &= 
    \begin{bmatrix}
               1 & 0 & 0 \\
           0 & \cos{\frac{\theta}{2}} & -e^{-i\phi}\sin{\frac{\theta}{2}} \\
           0 & e^{i\phi}\sin{\frac{\theta}{2}} & \cos{\frac{\theta}{2}}
    \end{bmatrix}.
\end{align}
\end{subequations}
\begin{figure}[h]
    \centering
    \includegraphics{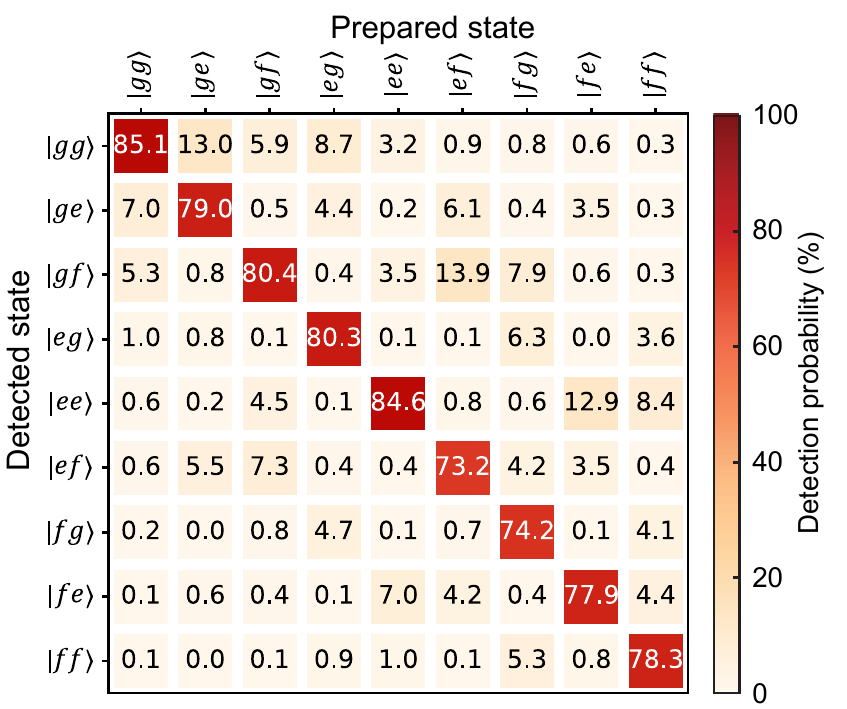}
    \caption{Single shot confusion matrix. Two-qutrit basis states are prepared and measured 5000 times.}
    \centering
    \label{fig:heatmap}
\end{figure}

Simultaneous single-shot readouts are collected after each of the 81 rotations. Fig.~\ref{fig:heatmap} shows the single shot confusion matrix of our readout. To compensate for the measurement error, we applied the inverse of the confusion matrix to the readout result. Maximum-Likelihood-Estimation (MLE) is used to reconstruct the physical density matrix $\rho_{m}$ that minimizes the cost function $f_c$,
\begin{equation}
\begin{split}
f_c(\overrightarrow{p},\overrightarrow{q})&=\sum_{j=1}^{81}\sum_{a,b=g,e,f}\left(\frac{p_{j,\ket{ab}}-q_{j,\ket{ab}}}{q_{j,\ket{ab}}}\right)^2 ,\\
p_{j,\ket{ab}} &= \bra{ab}S_{j}\cdot\rho_{m}\ket{ab}, \\
q_{j,\ket{ab}} &= \bra{ab}S_{j}\cdot\rho_{exp}\ket{ab}. \\ 
\end{split}
\end{equation}
Here $q_{j,\ket{ab}}$ is the measured probability for $\ket{ab}$ after the $j$th tomography rotation. For any state tomography data, we repeat the same experiment 5000 times to approximate each $q_{j,\ket{ab}}$.
We first obtain $\rho_{exp}$ from direct inversion of the experimental data and then perform MLE to find the physical density matrix $\rho_{m}$.

\begin{figure*}[t]
    \centering
    \includegraphics[width=\textwidth]{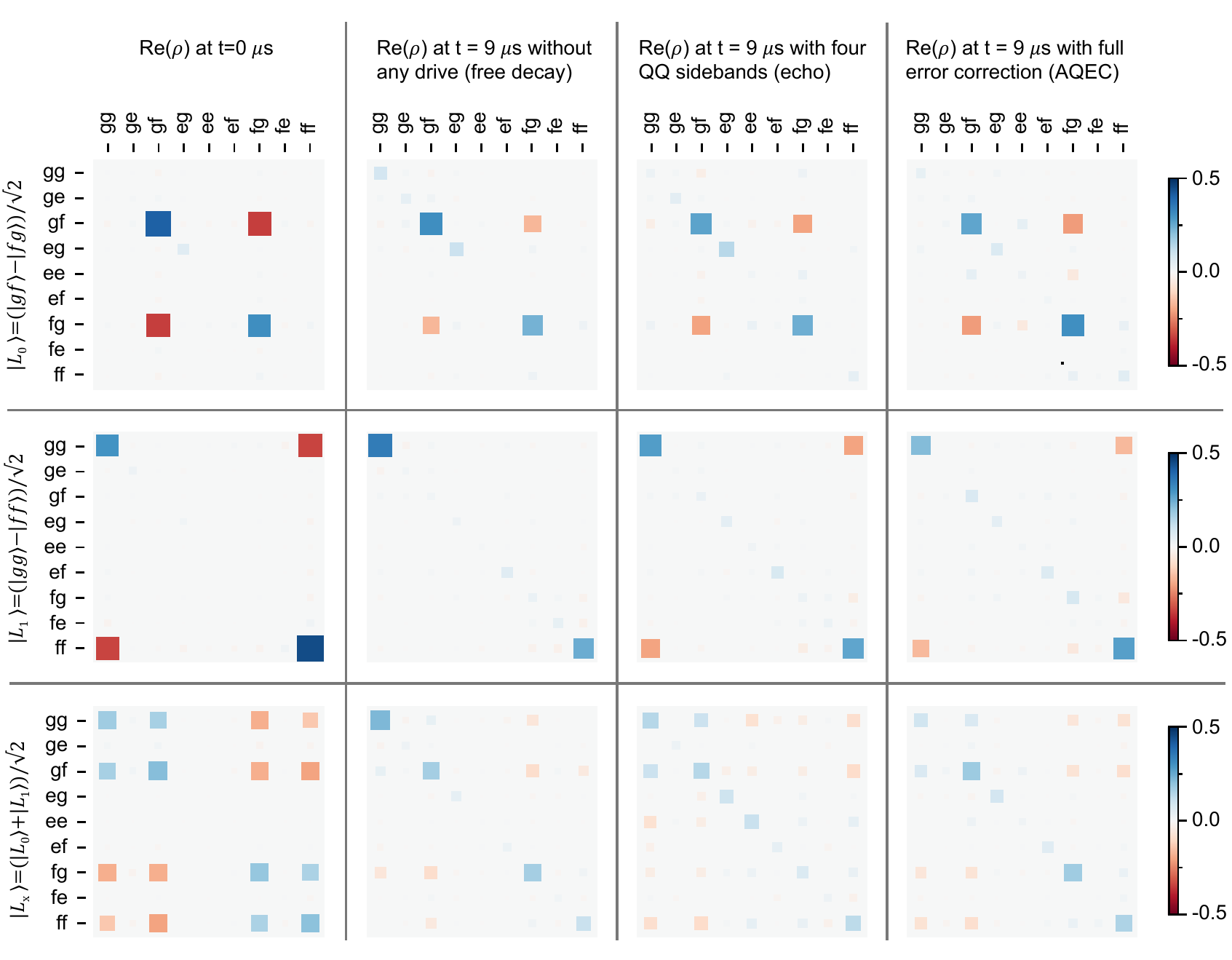}
    \caption{Evolution of the logical states under different conditions. Panels from top to bottom correspond to the logical state $\ket{L_0}$, $\ket{L_1}$ and $\ket{L_x}$. The real part of the density matrices are plotted as the imaginary components are small after phase rotation. The left column shows the initial states. Improvements in the coherence can be seen for the echo case when compared to free decay. With the full Star code protocol, further improvements are observed.}
    \centering 
    \label{fig:dm_evol}
\end{figure*}

\begin{figure}[t]
    \centering
    \includegraphics[width=\columnwidth]{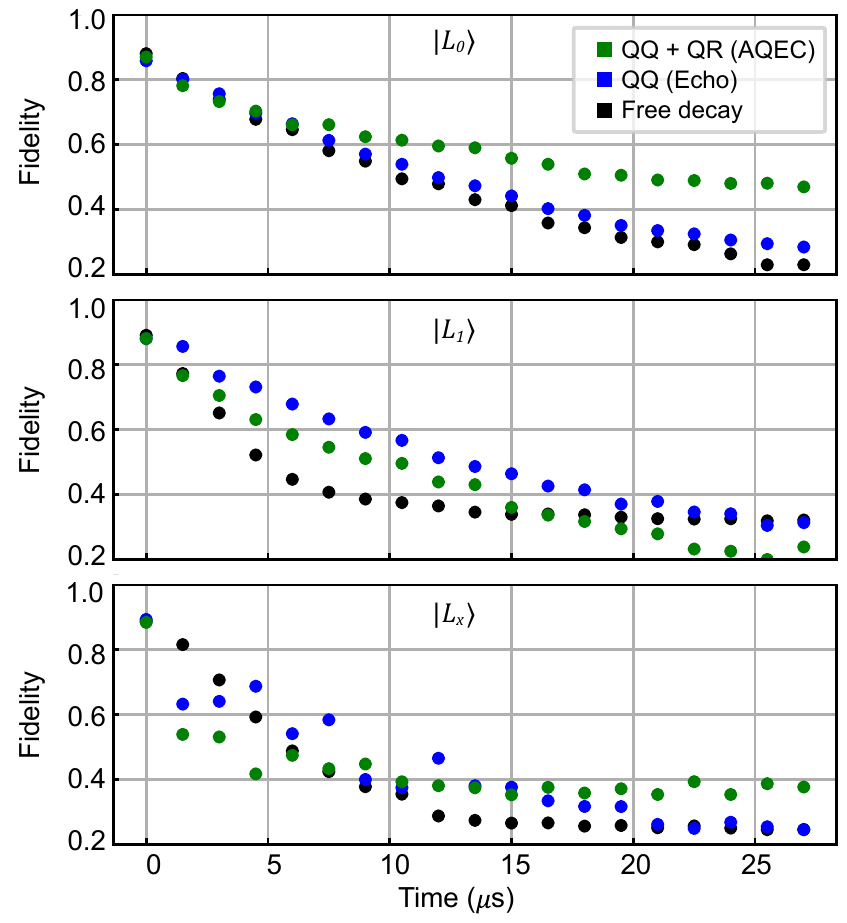}
    \caption{Experimental logical state fidelity as a function of time. Fidelities are calculated at each time point through state tomography. $\ket{L_1}$ has higher fidelity in the free decay case because of contribution from $\ket{gg}$.}
    \centering 
    \label{fig:fidelity_sweep}
\end{figure}

The tomographically reconstructed states after preparation, and after $\SI{9}{\micro\second}$ for the three cases of free decay, 4 QQ echo, and AQEC are illustrated in Fig.~\ref{fig:dm_evol}. The evolution of fidelities as a function of time are shown in Fig.~\ref{fig:fidelity_sweep}.
\section{Simulation and error channels in the AQEC}\label{app:simulation}
\begin{figure}[t]
    \centering    \includegraphics{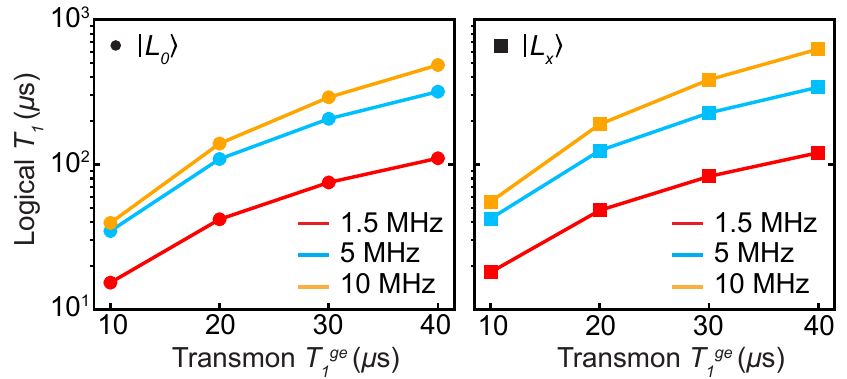}
    \caption{Theoretical logical lifetime in the rotating frame. The logical lifetime increases as a function of physical $T_{1}$. $\ket{L_1}$ has a similar lifetime as $\ket{L_0}$ in the simulation. A larger QQ sideband rate also provides higher logical qubit coherence. Parameter used for simulation are $\{W=10,\SI{5}{\mega\hertz}, \Omega_{j}=\SI{1.0}{\mega\hertz}, \nu_{r}=-\nu_{b}=\frac{W}{2}\}$ with $\{W=\SI{1.5}{\mega\hertz}, \Omega_{j}=\SI{0.4}{\mega\hertz}$, $\nu_{r}=-\nu_{b}=\SI{0.85}{\mega\hertz}\}$, $\kappa=\SI{0.5}{\mega\hertz}$.}
    \centering
    \label{fig:th_t1_with_t2}
\end{figure}

All simulations are carried out in a Hamiltonian of dimension $3\times3\times2\times2$. We first simulate the theoretical lifetime improvement with only photon loss error in the rotating frame (Eq.~\ref{eq:fully_rotated1}), and results are shown in Fig.~\ref{fig:th_t1_with_t2}. All simulated data show improvements beyond the break-even point, even with only $\SI{10}{\micro\second}$ $T_{1}^{ge}$ and modest rate requirements for QQ and QR sidebands. The logical coherence limits come from the double photon loss event and off-resonant population to other stray states from the spectrum crowding (see Ref.~\cite{VStar2023}). Logical errors will happen when a second photon decays before correction. $\ket{L_x}$ has a higher lifetime than $\ket{L_j}$ because it is partially protected against double photon loss in a single transmon. Longer physical $T_{1}^{ge}$, larger $W_{r/b}$, and $\Omega_{j}$ for a faster error correction rate help increase this limit. 

To simulate the real system, several error channels are introduced in the static frame $\tilde{H}_{\rm static}$ (Eq.~\ref{eq:logic_static})--- single photon decay $T_1^{ge,j}$ and $T_1^{ef,j}$, single transmon dephasing $T_{\phi}^{j}$, single-photon excitation $T_{1}^{\uparrow}$, resonator photon population $n_{\rm res}$ and extra correlated dephasing $T_{\phi}^{\ket{ff}}$ at $\ket{ff}$ level. Since only $ZZ_{ff1}$ and $ZZ_{ff2}$ have effects on the logical state, we model the $ZZ$s by directly adding energy shifts to $\ket{fe}$ and $\ket{ef}$, so that all logical basis still share the same energy and remain static in the frame. In the presence of external drives, the parameters will be different from the free decay case. We use experimentally measured $ZZ_{ff1}$ and $ZZ_{ff2}$ values in the simulation. The full master equation is solved in QuTip, 
\begin{equation}
\begin{split}
    \frac{\partial\rho(t)}{\partial t}&=-i\left[H_{\rm full}, \rho(t)\right]\\
    &+(\sum_{j=1,2}(\frac{1}{T_1^{ge,j}}D_j[\ket{g}\bra{e}]+\frac{1}{T_{1}^{ef,j}}D_j[\ket{e}\bra{f}]\\
    &+\frac{1}{T_1^{\uparrow}}D_j[\ket{e}\bra{g}]+\frac{2}{T_{1}^{\uparrow}}D_j[\ket{f}\bra{e}]\\
    &+\frac{1}{T_{\phi}^{j}}D_j[\ket{e}\bra{e}]+\frac{4}{T_{\phi}^{j}}D_j[\ket{f}\bra{f}]\\
     &+\kappa_{j}n_{\rm res}D[a_{rj}^{\dag}]+\kappa_{j}D[a_{rj}])\\
     &+\frac{1}{T_{\phi}^{\ket{ff}}}D_{12}[\ket{ff}\bra{ff}])\rho(t).
\end{split}
\end{equation}
Here we define
\begin{align}
H_{\rm full}&=\tilde{H}_{\rm static}+\sum_{j=1,2}\chi_{j}n_{qj}n_{rj}\nonumber\\
&+(ZZ_{ff1}\ket{fe}\bra{fe}+ZZ_{ff2}\ket{ef}\bra{ef})\otimes I_4,\\
D[A]\rho&=A\rho A^{\dag}-\frac{1}{2}\left(A^{\dag}A\rho+\rho A^{\dag}A\right)\nonumber,\\
D_1[A]&=D[A\otimes I_3\otimes I_4]\nonumber,\\
D_2[A]&=D[I_3\otimes A\otimes I_4]\nonumber,\\
D_{12}[A]&=D[A\otimes I_4]\nonumber.
\end{align}

Since transmons' anharmonicities are much larger than the transmon decay rate, each level's decay and dephasing are phase-independent. The system's full density matrix $\rho(t)$ is calculated and used to extract the coherence time and correctable error rate. Table~\ref{table:simulation_parameter} includes all parameters used in the master equation simulation. For each separate case (free decay, 4 QQ echo, and AQEC), parameters are the same for all logical states $\ket{L_0}$, $\ket{L_1}$, and $\ket{L_x}$. $T_{\phi}^{j}$ is increased in the 4 QQ Echo and AQEC cases because of the echo suppression of $1/f$ noise.

\begin{table}[t]
		\begin{tabular}{c c c c}
		    \hline
		    \hline
		     Simulation parameters & Free decay  & 4 QQ echo & AQEC\\  \hline
		    $Q_{1}\ T_{1}^{ge,1} (\mu s)$ & $18.0$ & $21.0$ & $21.0$ \\ 
		    $Q_{1}\ T_{1}^{ef,1} (\mu s)$ & $33.0$ & $29.0$ & $23.0$ \\ 
		    $Q_{1}\ T_{\phi}^{1} (\mu s)$$^{\dagger}$ & $15.0$ & $23.0$ & $23.0$ \\  
		    $Q_{1}\ T_{1}^{\uparrow} (\mu s)$ & $\infty$ & $\infty$ & $600.0$\\ 
		    \hline
		    $Q_{2}\ T_{1}^{ge,2} (\mu s)^{*}$ & $8.0$ & $9.0$ & $9.0$ \\ 
		    $Q_{2}\ T_{1}^{ef,2} (\mu s)$ & $33.0$ & $29.0$ & $23.0$ \\ 
		    $Q_{2}\ T_{\phi}^{2} (\mu s)$$^{\dagger}$ & $15.0$ & $23.0$  & $23.0$ \\  
		    $Q_{2}\ T_{1}^{\uparrow} (\mu s)$ & $\infty$ & $\infty$ & $600.0$\\  
		    \hline
		    $T_{\phi}^{\ket{ff}} (\mu s)$ & $4.4$ & $80.0$  & $80.0$\\ 
            $\kappa_1$ (MHz) & $0.53$ & $0.53$ & $0.53$ \\
            $\kappa_2$ (MHz) & $0.48$ & $0.48$ & $0.48$ \\
            $\chi_{1}$ (MHz) &  &  & $-0.2$ \\
		    $\chi_{2}$ (MHz) &  &  & $-0.2$\\
		    $n_{\rm res}$ & $0.00$ & $0.00$ & $0.03$ \\
		    \hline
		    $W_r$ (MHz) &  & $1.00$ & $1.45$\\
		    $W_b$ (MHz) &  & $1.70$ & $1.25$\\
		    $\nu_{r}$ (MHz) &  & $1.50$ & $0.80$\\
		    $\nu_{b}$ (MHz) &  & $0.00$ & $-0.90$\\
		    $\Omega_{1}$ (MHz) & & & $0.39$\\
		    $\Omega_{2}$ (MHz) & & & $0.39$\\
		    $ZZ_{ff1}$ (MHz) & & & $0.6$ \\
		    $ZZ_{ff2}$ (MHz) & & & $2.2$\\  
		\end{tabular}
		\caption{Parameters used in the master equation simulation. $\{W_{r/b}, \Omega_j, \nu_{r/b}\}$ are extracted through Fig.~\ref{fig:sweep_block_calibraion} in simulation; $ZZ_{ff1}$ and $ZZ_{ff2}$ are experimentally measured through $\ket{e0}\leftrightarrow\ket{f1}$ on-resonance frequency difference when all sidebands on. Coherence times and $\chi_{j}$ in the simulations are tuned to explain the Star code experiment, which are slightly different from the measurement. Irrelevant parameters in each case are not shown in the table and not included in the simulation. \\
        $^{\dagger}$ Dephasing in the 4 QQ echo and AQEC cases are higher because of the QQ sideband spin-echo improvement.\\
		$^{*}$ $Q_{2}$'s $T_{1}^{ge}$ is lower than $Q_{1}$'s because of the TLSs around the transition frequency. Effects to the codewords performance are minimal as population on $\ket{e}$ is corrected.\\}
	
		\label{table:simulation_parameter}
\end{table}
\begin{table*}[t]
		\begin{tabular}{c c c c}
		    \hline
		    \hline
		     Error channels limit & $\ket{L_0}$ & $\ket{L_1}$  &  $\ket{L_x}$ \\  \hline
		    Ideal implementation$^{*}$ & \multicolumn{2}{c}{$\SI{95}{\micro\second}$} & $\SI{160}{\micro\second}$ \\        
		    \hline
		    Transmon photon excitation & \multicolumn{2}{c}{$\sim\SI{360}{\micro\second}$} & $\sim\SI{3}{\milli\second}$ \\ 
		    $n_{\rm res}$ dephasing$^{\dagger}$ & \multicolumn{2}{c}{$\sim\SI{55}{\micro\second}$} & $\sim\SI{25}{\micro\second}$ \\  
		    Other dephasing noise$^{\dagger}$ & \multicolumn{2}{c}{$\sim\SI{50}{\micro\second}$} & $\sim\SI{25}{\micro\second}$\\
		    Transmon $ZZ$ dephasing & \multicolumn{2}{c}{$\infty$}  & \multirow{2}{*}{$\sim\SI{25}{\micro\second}$}\\ 
		    QR frequency mismatch & $>\SI{10}{\milli\second}$ & $\sim\SI{45}{\micro\second}$ \\
		    QQ frequency mismatch & \multicolumn{2}{c}{$>\SI{10}{\milli\second}$} & $\sim\SI{1.5}{\milli\second}$\\
		    QQ rate mismatch & \multicolumn{2}{c}{$>\SI{10}{\milli\second}$}  & $\sim\SI{4}{\milli\second}$ \\  
		    Reduced physical $T_1$ &\multicolumn{2}{c}{$\sim\SI{330}{\micro\second}$} &$\sim\SI{400}{\micro\second}$ \\
		    \hline
		    Experimental lifetime  & $23.4\pm\SI{6.8}{\micro\second}$ &$16.9\pm\SI{3.7}{\micro\second}$ &$8.7\pm\SI{4.6}{\micro\second}$ \\
		\end{tabular}
		\caption{Various decoherence channels for the logical qubit. Ideal implementation represents logical states' lifetime with QR coupling and only $T_{1}$ error. Each limit is extracted using the simulation through lifetime difference after adding relevant error channels. The average photon number in the resonator ($n_{\rm res}$) increases during external drives and dephases transmons through photon shot noise. Other dephasing noise include $1/f$ noise, white noise, correlated dephasing noise, and any other noise source. The total effect is represented with $T_{\phi}^{j}$ in the simulation. $ZZ$ between transmons introduces a large mismatch in QR frequency for $\ket{L_1}$ and $\ket{L_x}$, and the effect is combined with $ZZ$ dephasing for $\ket{L_x}$ case. The drifts in sideband amplitudes frequencies are less than $5\%$ and $\SI{20}{\kilo\hertz}$, and those limits are in the order of $\SI{}{\milli\second}$.\\
        $^{*}$ Ideal implementation includes QR couplings $\chi_j$ but no QQ $ZZ$ couplings.\\
		$^{\dagger}$ $n_{\rm res}$ and $T_{\phi}^{j}$ are determined through simulation-experiment matching.}
		\label{table:error_channel}
\end{table*}

Table~\ref{table:error_channel} shows the lifetime limitations from different error channels in the AQEC case. In the ideal implementation, we include only the single photon decay and QR couplings $\chi_j$ in the simulation. The transmon photon excitation is enhanced when all sidebands are turned on. However, excitation error on $\ket{L_x}$ is partially correctable under three-level approximation, and thus $\ket{L_x}$ is more insensitive to it compared to $\ket{L_{0/1}}$. Resonator photon excitation happens from the heating effect when QR sidebands are on. Larger cavity photon number $n_{\rm res}$ will dephase all logical states and is one of the dominant error sources in our system. With higher resonator frequencies or an extra coupler between transmon and resonator, $n_{\rm res}$ can be reduced under the same QR rate $\Omega_{j}$. The next two dominant error channels are cross-Kerr of between the transmons and QR frequency mismatch. $ZZ_{ff1}$ and $ZZ_{ff2}$ will dephase logical superposition states, as discussed in Appendix~\ref{section:ZZ} (but has no effect on individual logical states). QR frequency mismatch is unavoidable in the presence of $ZZ$. In the experiment, we apply on-resonance $\ket{e0}\leftrightarrow\ket{f1}$ drive for $\ket{L_0}$ (corresponding the partner transmon being in $\ket{g}$). Consequently, for $\ket{L_1}$ the QR sidebands become detuned by $ZZ_{ff1}$ and $ZZ_{ff2}$ (corresponding to the partner being in $\ket{f}$) and effectively perform slower error correction. The QQ sideband frequency mismatch comes from a modest upper bound of the system's frequency drift (around $\SI{10}{\kilo\hertz}$). This is not comparable to the $W$ and has no significant influence on the logical states. Other dephasing noise sources include $1/f$ noise, white noise, and $\ket{ff}$'s correlated dephasing. Among those three the white noise affects AQEC performance as it has a constant noise spectrum that cannot be suppressed through the spin echo. Star code protocol is also insensitive to small sideband amplitude drifts. The phase between logical states is defined by different sideband pairs, and amplitude drifts have to be comparable to $W$ to change the logical states. Further, when all sidebands are on, both transmons' physical $T_{1}$ are shortened, which slightly reduces the performance.
Other insignificant error sources include leakage to higher transmon energy levels ($|\alpha_{j}|\gg W$) and population in the coupler mode ($\omega_{c}\gg\omega_{j}$). Those are not considered in the simulation as the transition frequency is far away.

\section{Device Fabrication and Measurement Setup}
\begin{figure*}[t]
    \centering
    \includegraphics{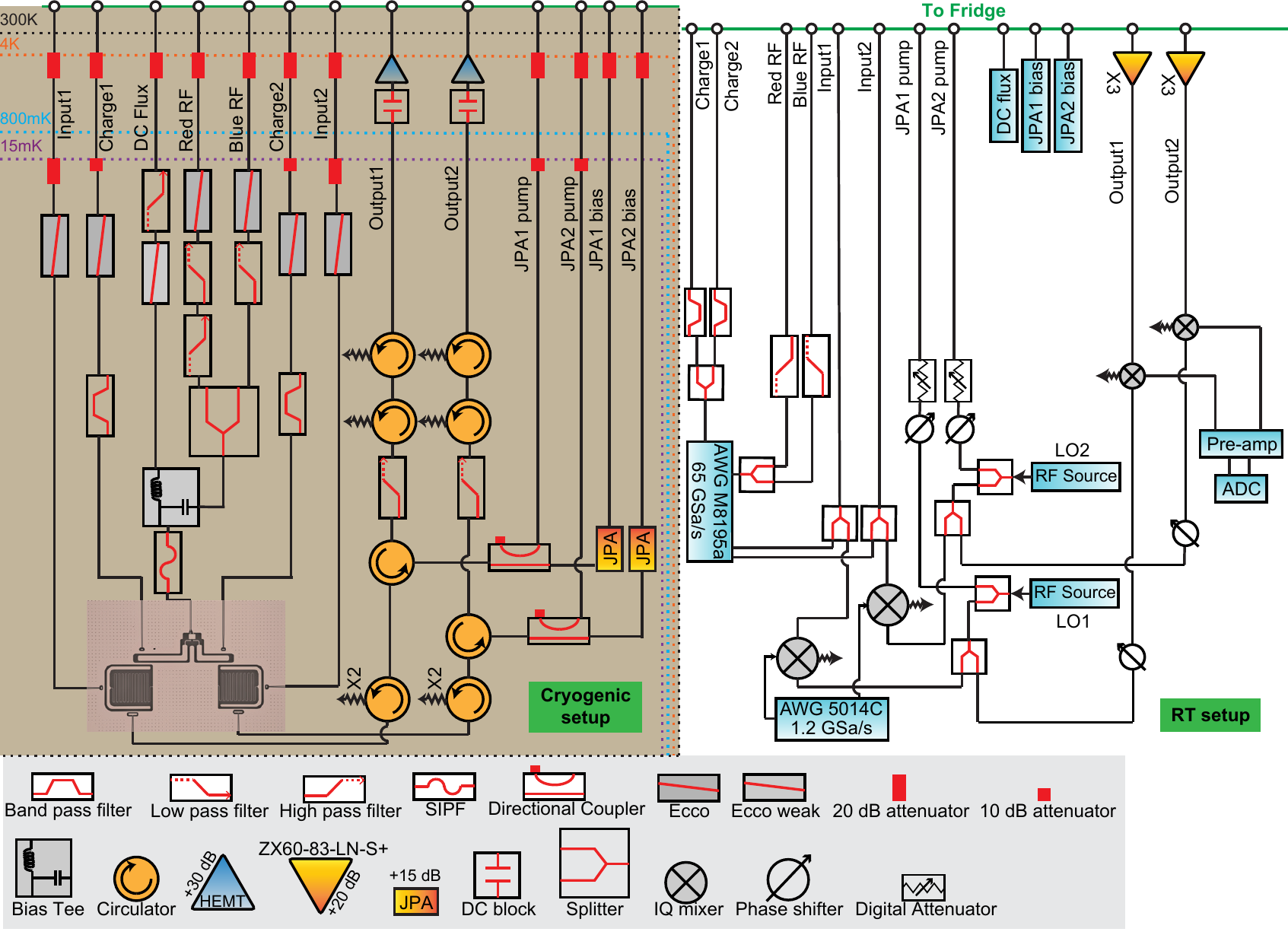}
    \caption{Detailed cryogenic and room temperature measurement setup.}
    \centering
    \label{fig:measurement}
\end{figure*}

The substrate for the device is a 430~$\mu$m thick C-plane sapphire wafer annealed at $1200^o$C for 2 hours. The ground plane uses 200~nm thick Tantalum film sputtered at $800^o$C. Large patterns, except Josephson junctions, were made through optical lithography and 20-second wet-etching in Transene Tantalum etchant 111. AZ 1518 was spin-coated as the positive photoresist, and a Heidelberg MLA 150 Direct Writer was used for the photolithography. The junction mask was fabricated with a Raith EBPG5000 Plus E-Beam Writer on a bi-layer resist (MMA EL11-950 PMMA A7). Transmon and coupler's Josephson junctions are Dolan bridge type. The mask was evaporated in a Plassys electron-beam evaporator with double-angle evaporation ($\pm 23^o$). The wafer was diced into $7\times7$ mm$^2$ chips and lifted off. After measuring the test junctions' resistances, the device was mounted on a printed circuit board, wire-bonded, packaged inside a $\mu$-metal shielded sample can, and installed inside a dilution fridge. 

Figure~\ref{fig:measurement} shows the room and cryogenic temperature measurement chain. The device is mounted on the mixing chamber plate of the dilution fridge with a 15 mK base temperature. A Tektronix 5014C AWG (1.2GSa/s) acts as the master trigger for all other equipment. The readout pulses are generated through two CW tones from RF sources (PSG-E8257D), modulated by AWG 5014C. The qubit input pulses are generated through another 4-channel AWG (Keysight M8195 65 Gsa/s, 16 Gsa/s per channel). The qubit and readout signals are combined and sent through lines In$_{1}$ and In$_{2}$ into the dilution fridge. Three DC sources (Yokogawa GS200) are used to bias the DC flux of the coupler and two Josephson Parametric Amplifiers (JPA). The red and blue QQ RF flux drives and two direct QR charge drives are synthesized through the same 4-channel AWG. Inside the fridge, at the 4K plate, all input lines have 20-dB attenuators. At the base plate, In$_{1}$ and In$_{2}$ lines have 10-dB attenuators, followed by a strong Eccosorb\textsuperscript{\textregistered} providing 20-dB attenuation at $\SI{4}{\giga\hertz}$. Charge$_{1}$ and Charge$_{2}$ lines have 20-dB attenuators, followed by strong Eccosorb providing 20-dB attenuation at $\SI{4}{\giga\hertz}$, and a bandpass filter with passband $3.9 - \SI{4.8}{\giga\hertz}$. The DC Flux line has a low pass filter (DC $-~\SI{1.9}{\mega\hertz}$) and a weak Eccosorb (1-dB attenuation at $\SI{4}{\giga\hertz}$). The red-frequency RF flux line passes through a weak Eccosorb first, followed by a high pass filter (cut off at $\SI{200}{\mega\hertz}$) and a low pass filter (cut off at $\SI{2}{\giga\hertz}$). The blue-frequency RF flux line passes through a weak Eccosorb first, followed by a high pass filter (cut off at $\SI{6}{\giga\hertz}$). The two RF flux lines and the DC flux line are combined and pass through a Step Impedance Purcell Filter (SIPF) with a stop band $2.5 - \SI{5.5}{\giga\hertz}$. The two output signals go through three circulators, then each amplified by a JPA with 15-dB gain, followed by a low pass filter (cut off at $\SI{8}{\giga\hertz}$), two circulators, a DC block, and amplified with one LNF High-Electron-Mobility Transistor (HEMT) amplifier. The output signals are further amplified at room temperature, then demodulated, filtered with a low pass filter (DC $- \SI{250}{\mega\hertz}$), and amplified again using the SRS Preamplifier. The final signal is digitized with Alazar ATS 9870 (1GSa/s) and analyzed in a computer. 

\clearpage
\normalem{}
\bibliography{2Q}
\end{document}